\documentclass[11pt]{article}

\usepackage{amsmath,amssymb}
\usepackage{slashed}
\usepackage{xcolor} 
\usepackage{graphicx}
\usepackage{url}

\newcommand{\ttbar}{{t\bar t}}
\newcommand{\Al}{{A_{\ell}}}
\newcommand{\ACl}{{A_{C}^{\ell}}}
\newcommand{\AClt}{{A_{C}^{t \ell}}}
\newcommand{\Attb}{{A_{t \bar t}}} 
\newcommand{\ACttb}{{A_{C}^{t \bar t}}}   
\newcommand{\mttb}{{m_{t \bar t}}} 
\newcommand{\ptl}{{p_{T,\ell}}} 

\newcommand{\ttb}{\ensuremath{{t\bar{t}}}}
\newcommand{\ACll}{{A_{C}^{\ell \ell}}}
\newcommand{\All}{{A_{\ell \ell}}}

\newcommand{\beq}{\begin{equation}}
\newcommand{\eeq}{\end{equation}}
\newcommand{\bea}{\begin{eqnarray}}
\newcommand{\eea}{\end{eqnarray}}
\newcommand{\nn}{\nonumber \\ }

\newcommand{\tev}{{\mathrm TeV}}
\newcommand{\hc}{\mathrm{h.c.}}

\begin{document}

\vspace*{-2cm}
\begin{flushright}
CERN-PH-TH/2014-005  \\
IPM/P.A-333  \\ 
LPT Orsay 14-05 \\
\vspace*{2mm}
\today
\end{flushright}

\begin{center}
\vspace*{15mm}

\vspace{1cm}
{\Large \bf 
From Tevatron's top and lepton-based asymmetries to the LHC  
} \\
\vspace{1cm}

{\bf Adri\'an Carmona${}^1$,  Mikael Chala${}^2$, Adam Falkowski${}^3$, Sara Khatibi${}^4$, \\
 Mojtaba Mohammadi Najafabadi${}^4$, Gilad Perez${}^{5,6}$, Jos\'e Santiago${}^2$}

 \vspace*{.5cm} 
  {\small\sl$^1\,$ Institute for Theoretical Physics, ETH Zurich, 8093 Zurich, Switzerland } \\
   {\small\sl$^2\,$ CAFPE and Departamento de F\'isica Te\'orica y del Cosmos, Universidad de Granada, E-18071 Granada, Spain } \\
 {\small\sl$^3\,$ Laboratoire de Physique Th\'eorique, CNRS -- UMR 8627, 
Universit\'e de Paris-Sud 11, F-91405 Orsay Cedex, France} \\ 
 {\small\sl$^4\,$ School of Particles and Accelerators, Institute for Research in Fundamental Sciences (IPM) P.O. Box 19395-5531, Tehran, Iran} \\
 {\small\sl$^5\,$Department of Particle Physics and Astrophysics, Weizmann Institute of Science, Rehovot 76100, Israel} \\
{\small\sl$^6\,$PH-TH Department, CERN, CH-1211 Geneva 23, Switzerland}

\vspace*{.2cm}

\end{center}

\vspace*{10mm}
\begin{abstract}
We define a lepton-based asymmetry in semi-leptonic $t\bar{t}$ production at the LHC. 
We show that the ratio of this lepton-based asymmetry and the $t\bar{t}$ charge asymmetry, measured as a function
of the lepton transverse momentum or the $t\bar{t}$ invariant mass is a robust observable in the Standard Model. 
It is stable against higher order corrections and mis-modeling effects. 
We  show that this ratio can also be a powerful discriminant among different new physics models and between them and
the Standard Model. Finally, we show that a related ratio defined at the Tevatron is also robust as a function of the $t\bar{t}$ invariant mass. 
\end{abstract}

\vspace*{3mm}

\section{Introduction} \label{sec:intro}

The top is unique among the known elementary fermions, it has several properties making it an object worth studying. 
From the experimental perspective, its complex structure provides many handles that are translated to a very rich set of observables to probe. From the perturbative QCD side the top is an object that enable theorists to make precise computations that yield accurate predictions to test against data. Within the Standard Model (SM), the top quark is also linked to flavor and electroweak physics due to its large Yukawa coupling. In fact, despite being perturbative, the sizable top Yukawa coupling implies that the top interactions at the quantum level dominate many of the flavor violating observables as well as the contributions to various electroweak observables. 
These features by themselves provide a fairly good motivation to transform top physics into a generic sensitive tool for new physics searches. However, what really singles out top physics as a major player in the new physics searches frontier is the fine-tuning problem. We now know for fact that the Higgs boson exists, and it happens to be pretty light. We are also reasonably certain that the SM Higgs mechanism, with its fundamental scalar, plays a dominant role in electroweak symmetry breaking. This implies that the Higgs mass is subject to large quantum corrections. The largest corrections are induced by the top-Higgs couplings. As is well known, the only well established and concrete mechanism to solve this UV sensitivity problem is to extend the top sector to include new light ``top-partners" that would counterbalance the top quantum corrections to the Higgs mass. Thus, studying top physics is expected to shed light about the mechanism of electroweak scale stabilization. It transforms top physics into a window for new physics searches with a rough scale associated to them that is expected to be within the LHC reach.

Not all precision top observables provide a direct link with the physics of naturalness. An example for such an observable is the top pair forward-backward asymmetry (and its derivatives to be discussed in the following). The reasoning behind this statement is the fact that, to generate a sizable asymmetry, one requires the new dynamics to have a sizable coupling to the tops as well as the first generation quarks, the proton-anti-proton valence constituents. 
As the fine-tuning ``pressure" coming from the light quarks is negligible it is hard to make a case for a direct linkage between this observable and natural models of electroweak symmetry breaking.\footnote{One can find exotic examples, though, in which an indirect linkage is found between the physics of electroweak symmetry breaking and naturalness and potentially large $t\bar t$ forward-backward asymmetries. Such an example could be related to composite Higgs models in the presence of composite first two generation quarks (in addition to the top ones) which are allowed by precision observables~\cite{Carena:2007ua,Atre:2008iu,flavor-triviality1,Redi:2011zi}, lead to a sizable deviation in the Higgs couplings~\cite{Atre:2013ap,Delaunay:2013iia} and potentially also to a sizable~\cite{flavor-triviality1,DaRold:2012sz} asymmetry.}

In this paper we consider a set of $t\bar t$ asymmetries, where our starting point is related to the Tevatron anomalous forward--backward asymmetry.
Within the SM, the $\ttbar$\/ forward--backward asymmetry, $\Attb$,
is an interesting variable because it tells us about QCD interactions
beyond leading order but in a region that should be well described by
perturbation theory~\cite{Kuhn:1998kw,Kuhn:1998jr}. Furthermore, as
the SM contributions are expected to be
small~\cite{Kuhn:1998kw,Kuhn:1998jr,Bowen:2005ap,Antunano:2007da,Almeida:2008ug},
the measurement of $\Attb$ is sensitive to beyond-the-SM (BSM)
contributions. As mentioned, the asymmetry is quite of a special observable since
shifting it requires new physics with non-standard couplings both to
the $t\bar t$\/ quark current as well as to the current of
$u\bar u$\/ (or possibly $d\bar d\,$) initial-state
quarks.

 Both Tevatron
experiments, CDF and D$\O$, have observed an anomalously large
forward-backward asymmetry in $\ttb$ production, defined by
\begin{equation}
\Attb=\frac{N(\Delta y^{t\bar{t}}>0)-N(\Delta y^{t\bar{t}}<0)}
{N(\Delta y^{t\bar{t}}>0)+N(\Delta y^{t\bar{t}}<0)},
\end{equation}
where $\Delta y^{t\bar{t}}\equiv y_t-y_{\bar{t}}$ and $N$ is the
total number of events
satisfying the corresponding constraint. This asymmetry has been
measured in semi-leptonic decays with the following result: 
\begin{eqnarray}
\Attb({\rm CDF}) &=& 0.164\pm 0.047,\quad \cite{Aaltonen:2012it}, \\
\Attb ({\rm D\O}) &=& 0.196\pm 0.065,\quad \cite{Abazov:2011rq},
\end{eqnarray}
to be compared with the SM NLO prediction with electroweak corrections
included~\cite{Bernreuther:2012sx},  
\begin{equation}
\Attb(\mathrm{SM})= 0.088\pm 0.006.
\end{equation}
Although not statistically significant for a discovery, the observed excess  is consistent among experiments. 
Moreover, the excess in the top asymmetry  is accompanied by several excesses in  lepton-based asymmetries measured at the Tevatron  in the semi-leptonic (SL) and di-leptonic (DL) channels. 
The current results for inclusive lepton-based asymmetries  together with the SM prediction (as reported by the experimental collaborations) are  
\begin{equation}
\begin{array}{llll}
{\rm SL:} &\Al({\rm CDF})= 0.094 \pm 0.024^{+0.022}_{-0.017}, \quad
&  \Al ({\rm SM})= 0.038\pm 0.003,  & \cite{Aaltonen:2013vaf}, 
\\
{\rm SL:} & \Al({\rm D\O}) = 0.047\pm 0.023^{+0.011}_{-0.014},\quad 
&\Al ({\rm SM}) = 0.023,\quad & \cite{D0:Alsemi-leptonic}, 
\quad
\\
{\rm DL:} & \Al ({\rm CDF}) = 0.072\pm 0.052 \pm 0.030, \quad
\quad
&\Al ({\rm SM})= 0.038 \pm 0.003  ,\quad  & \cite{cdfall}, 
\\
{\rm DL:} & \Al ({\rm D\O}) = 0.044\pm 0.037\pm 0.011,\quad 
\quad
&\Al ({\rm SM})= 0.024 \pm 0.001,\quad  & \cite{Abazov:2013wxa},
\\
{\rm DL:} & \All ({\rm CDF}) = 0.076 \pm 0.072 \pm 0.037,\quad 
\quad
&\All ({\rm SM}) = 0.048 \pm 0.004,\quad &  \cite{cdfall},  
\\ 
{\rm DL:} & \All ({\rm D\O }) = 0.123\pm 0.054\pm 0.015,\quad 
\quad
&\All ({\rm SM}) = 0.048\pm 0.004,\quad & \cite{Abazov:2013wxa},
\end{array}
\end{equation}
where the single and double lepton-based asymmetries are defined as
follows
\begin{equation}
\Al = \frac{N(q\times \eta >0) -N(q\times \eta <0) }
{N(q\times \eta >0) +N(q\times \eta <0) }, \label{eq:Al:def}
\end{equation}
and
\begin{equation}
\All=\frac{N(\Delta \eta >0)-N(\Delta \eta <0)}
{N(\Delta \eta >0)-N(\Delta \eta <0)},
\end{equation}
with $q$ and $\eta$ the charge and pseudorapidity of the lepton
and $\Delta \eta\equiv\eta_{l^+}-\eta_{l^-}$.

A puzzling  aspect of the observed excess is that the large value of the
measured asymmetries are not accompanied by any sizable deviation in
other top observables, such as the total or differential $\ttb$ 
production cross sections. This strongly
constrains possible explanations of the anomalous forward-backward
asymmetry. 
An unfortunate obstacle for a satisfactory understanding of this anomaly is the fact that the Tevatron ceased its operations in 2011. 
With most of the data already analyzed new insight into  the asymmetry can only come from a new smart choice of observables,
 or from exploring the larger dataset of the Large Hadron Collider (LHC) data. 

In~\cite{Falkowski:2012cu} it was shown
that the study of the correlation of $\Attb$ with a lepton-based
asymmetry $\Al$, measured as a function of some kinematical
variable, such as the lepton $p_T$ 
can be a powerful discriminating observable from the following three reasons: 

The {\it first} is that the lepton-based asymmetry is simpler to measure just because of the fact that the lepton momenta are measured directly and the relevant corrections due to detector effects are rather small.

The {\it second} is that within the SM the correlation between the
$\ttbar$\/ forward--backward asymmetry $\Attb$ and the
corresponding lepton-based asymmetry $\Al$ -- at the differential
level -- is strong and rather clean theoretically~\cite{Falkowski:2012cu}. 
The correlation is easy to understand qualitatively, it stems from a combination of the vector nature of QCD (or the absence of polarization in the top production and decay) and the fact that the leading order corrections to the lepton kinematics are screened away due to the narrow width of the top. 
Hence a combined measurement of the two distributions
as a function of the lepton $p_T$ would lead to a potentially
unbiased and normalization-free test of the SM prediction. In~\cite{Falkowski:2012cu} the robustness of this correlation was successfully tested given various deformation of the SM distributions, namely scale dependence, the transverse momentum of the $t\bar t$ system and higher order effects in the decay and showering.  

The {\it third} is that beyond the SM this correlation is generically lost. The lepton
asymmetry is sensitive to different aspects of the interaction
depending on the kinematical regime. In particular, it depends on the polarization
(and therefore chirality) of initial-state  quarks near the $\ttb$ production threshold,  whereas it depends 
on the top kinematics and polarization at large values of the $\ttb$ invariant mass  \cite{Falkowski:2011zr,Berger:2011pu,Baumgart:2013yra, Agashe:2006hk}. 
Some of these aspects can be very different in the SM  and in models of new physics explaining the anomalous $\Attb$. 
For instance, near threshold the lepton-based asymmetry could arise due to a different  contribution of left- and right-handed initial-state quarks to the $t \bar t$ production, 
 as opposed to the unpolarized initial state in the SM. At large invariant $t\bar t$ masses the lepton asymmetry may be stronger (weaker) if the new physics dominantly couples to right handed (left handed) tops.
 A simple variable like the lepton $p_T$ can be used to
interpolate between the different kinematical regimes and display in
this way the sensitivity to the different ingredients generating the
asymmetry~\cite{Falkowski:2012cu}. 

A definite confirmation of the origin of the anomalous $\Attb$ might come from the larger $t \bar t$ dataset collected at the LHC. 
 It is important to emphasize, though, that even within the SM the
Tevatron and LHC observables differ in nature. In particular, the
dominant $\ttbar$\/ production mechanism and the kinematical reaches
available to the top quarks are clearly very different at the two
colliders; the Tevatron collides charge-asymmetric beams and top quark
production is dominated by quark--antiquark annihilation, while, at the
LHC, collisions are charge symmetric and top pair production is driven
by gluon--gluon collisions. Furthermore, non-SM dynamics can naturally
induce a large deviation for the forward--backward asymmetry at the
Tevatron without affecting the charge asymmetry at the
LHC~\cite{Drobnak:2012cz,Drobnak:2012rb,AguilarSaavedra:2012va,Alvarez:2012ca}.
Thus, another byproduct of our study below is to investigate whether at the LHC the lepton-based asymmetry can break this degeneracy in theory space, namely to be sensitive to the presence of new physics that explain the Tevatron anomaly in models where the charge asymmetry at the LHC is close to the SM prediction. 

The related
charge asymmetry $\ACttb$ in $t\bar{t}$ production is dwarfed by the
dominating symmetric contribution from initial-state gluon production
and although current measurements do not show any deviation from the
SM prediction, the large errors leave room for an anomalous
contribution. In this situation it is also pressing to investigate
alternative observables that allow us to obtain as much information as
possible from
current data. 

The main goal of this article is to extend the studies
in~\cite{Falkowski:2012cu} to LHC observables. For the sake of
concreteness we will focus on the semi-leptonic decay mode in which one
top decays hadronically and the other decays leptonically. 
We will define a new lepton-based asymmetry and study the correlation
between this asymmetry and $\ACttb$ 
as a function of the lepton transverse momentum $\ptl$  and the $t\bar{t}$ pair invariant mass $\mttb$. 
We will show that this new observable is robust at the LHC in the
SM. We will then consider a number of new physics models that
reproduce the Tevatron asymmetries while being compatible with all
other experimental data. The first class of models generate the asymmetry
by the s-channel exchange of a massive color octet vector
resonance (axigluon) with different chirality structure for its
couplings and different mass range. 
Another model we study here is one in which 
the asymmetry is induced by the t-channel exchange
of a complex $Z^\prime$ boson. The different chirality structures and
kinematics induced in these models can be disentangled by means
of the ratio of asymmetries measured as a function of the lepton $p_T$
or the $t \bar{t}$ invariant mass. Our studies are based on the LHC
run at $\sqrt{s}=8$ TeV. Nevertheless we expect these observables to
be particularly useful during the longer run at the upgraded LHC with
$\sqrt{s}=13$ TeV as a unique tool to fully explore the origin of the
anomalous forward-backward asymmetry.

The rest of the article is organized as follows. We describe the
current status of measurements of the $\ttb$ charge asymmetry $\ACttb$ and
associated di-lepton-based asymmetry $\ACll$ at the LHC in
section~\ref{sec:review}, in which we also introduce our new
lepton-based asymmetry, $\AClt$. The behaviour in the SM of $\ACttb$ and
$\AClt$ as a function of $\ptl$ and $\mttb$ and the robustness of the
ratio $\AClt/\ACttb$ measured as a function of these kinematical
variables are described in section~\ref{sec:sm}. We describe in
section~\ref{sec:bsm} our new
physics models, current constraints, and the potential of the ratio of
asymmetries as a function of $\ptl$ or $\mttb$ to discriminate among
them and with respect to the SM and we present our conclusions in
section~\ref{sec:conclusions}. We present in an Appendix a test of the
robustness of the ratio of lepton-based and forward-backward
asymmetries measured at the Tevatron as a function of the $\ttb$
invariant mass and provide a comparison of the $\ptl$ dependence of a
lepton-based asymmetry measured by D$\O$ and the SM prediction.

\section{Top Asymmetries at the LHC} \label{sec:review}
 
The LHC cannot generate a forward-backward asymmetry in $t\bar{t}$ production because the $pp$ initial state is symmetric. 
However, the different parton distribution functions of quarks and anti-quarks inside the proton make it possible for the top and anti-top rapidity distributions to be different.  
Therefore one can define a non-vanishing charge asymmetry,  
\begin{eqnarray}
\ACttb=\frac{N(\Delta|y|^{t\bar{t}}>0)-N(\Delta|y|^{t\bar{t}}<0)}
{N(\Delta|y|^{t\bar{t}}>0)+N(\Delta|y|^{t\bar{t}}<0)},
\end{eqnarray}
where $\Delta|y|^{t\bar{t}}\equiv |y_t|-|y_{\bar{t}}|$. 
Due to the dominant symmetric contribution from initial state gluons the SM predicts a small charge asymmetry,   
$\ACttb({\rm SM})= 0.0123 \pm 0.0005$ for $\sqrt{s} = 7$~TeV LHC and $\ACttb({\rm SM})= 0.0111 \pm 0.0004$  for $\sqrt{s} = 8$~TeV LHC  \cite{Bernreuther:2012sx}. 
In the semi-leptonic channel ATLAS and CMS find the following
(unfolded) values:  
\begin{eqnarray}
\ACttb({\rm ATLAS, 7~\tev}) =& 0.006\pm 0.010,\quad &
\cite{A_C:semi-leptonic:atlas},  
\nonumber \\ 
 \ACttb({\rm CMS, 7~\tev})  =&   0.004 \pm 0.010 \pm  0.011  &
 \cite{Chatrchyan:2012cxa}, 
 \nonumber   \\ 
\ACttb({\rm CMS, 8~\tev})  =& 0.005\pm 0.007\pm 0.006,\quad & \cite{A_C:semi-leptonic:cms} ,
\end{eqnarray}
while in the di-leptonic channel the measured values are 
\begin{eqnarray}
\ACttb({\rm ATLAS, 7~\tev})  =& 0.057\pm 0.024\pm 0.015,\quad & \cite{A_C:leptonic:atlas}, \nonumber \\ 
\ACttb({\rm CMS, 7~\tev}) =& 0.050\pm 0.043^{+0.010}_{-0.039},\quad & \cite{A_C:leptonic:cms}.
\end{eqnarray}
A related leptonic asymmetry can be defined in events in which both tops decay leptonically,  
\begin{eqnarray}
\ACll=
\frac{N(\Delta|\eta|^{l^+ l^-}>0)-N(\Delta|\eta|^{l^+
    l^-}<0)}
{N(\Delta|\eta|^{l^+ l^-}>0)+N(\Delta|\eta|^{l^+ l^-}<0)},
\end{eqnarray}
where $\Delta|\eta|^{l^+l^-}\equiv |\eta_{l^+}|-|\eta_{l^-}|$. 
This observable was  measured by ATLAS and CMS
\begin{eqnarray}
\ACll({\rm ATLAS, 7~\tev})  &=& 0.023\pm 0.012\pm 0.008,\quad\cite{A_C:leptonic:atlas}, \\ 
\ACll({\rm CMS, 7~\tev}) &=& 0.010\pm 0.015\pm 0.006,\quad\cite{A_C:leptonic:cms}, 
\end{eqnarray}
where the SM prediction is quoted as $\ACll({\rm SM}) = 0.004$.

Our goal is to define a new lepton-based asymmetry in semi-leptonic
$\ttb$ events that maintains the interesting properties of the
lepton-based asymmetries at the Tevatron, namely a unique
and robust discriminating power when correlated with the charge
asymmetry as a function of $\ptl$ or $\mttb$.
The following lepton-based asymmetry fulfills the requirements: 
\begin{equation}
\AClt=
\frac{N(\Delta|y|^{tl}>0)-N(\Delta|y|^{tl}<0)}
{N(\Delta|y|^{tl}>0)+N(\Delta|y|^{tl}<0)},
\end{equation}
where we define 
\begin{equation}
\Delta|y|^{tl}\equiv
\left\{ \begin{array}{ll}
|y_{l^{+}}|-|y_{\bar{t}}|, & \mbox{ for leptonic top decays} \\ 
|y_{t}|-|y_{l^{-}}|,
& \mbox{ for leptonic anti-top decays}.
\end{array}\right.
\end{equation}
It is clear that at large $\ptl$ or $\mttb$ the
lepton will inherit the top properties it decayed from and this
asymmetry will approach $\ACttb$. 
At smaller values, however, it will
become sensitive to other features like the polarization of the
initial quarks and can therefore show deviations between the SM and
new physics models.\footnote{
We have also considered another lepton-based asymmetry observable, 
$\ACl=\frac{N_{+}-N_{-}}{N_{+}+N_{-}}$, 
where $N_{\pm}=\int |\eta|  N_{l^\pm}({\eta})$ is the cumulative
number of events with the corresponding charged lepton weighted with
the absolute value of the lepton rapidity. This second asymmetry has
the advantage that it does not require full reconstruction but
unfortunately the $\ACl/\ACttb$ ratio turns out not to be robust and
we will disregard it in the following.}

\section{Charge and lepton-based asymmetries in the SM: distributions and robustness tests} \label{sec:sm}

In this section we are going to describe the behavior of the asymmetries defined above, $\ACttb$ and $\AClt$, as a function of
$\ptl$ and $\mttb$ in the SM. 
We will then proceed to analyze the robustness of
the ratio $\ACttb/\AClt$ measured as a function of these variables
against various reconstruction and simulation effects.

\begin{figure}[hbtp]
\centering
\includegraphics[width=7cm,height=5cm]{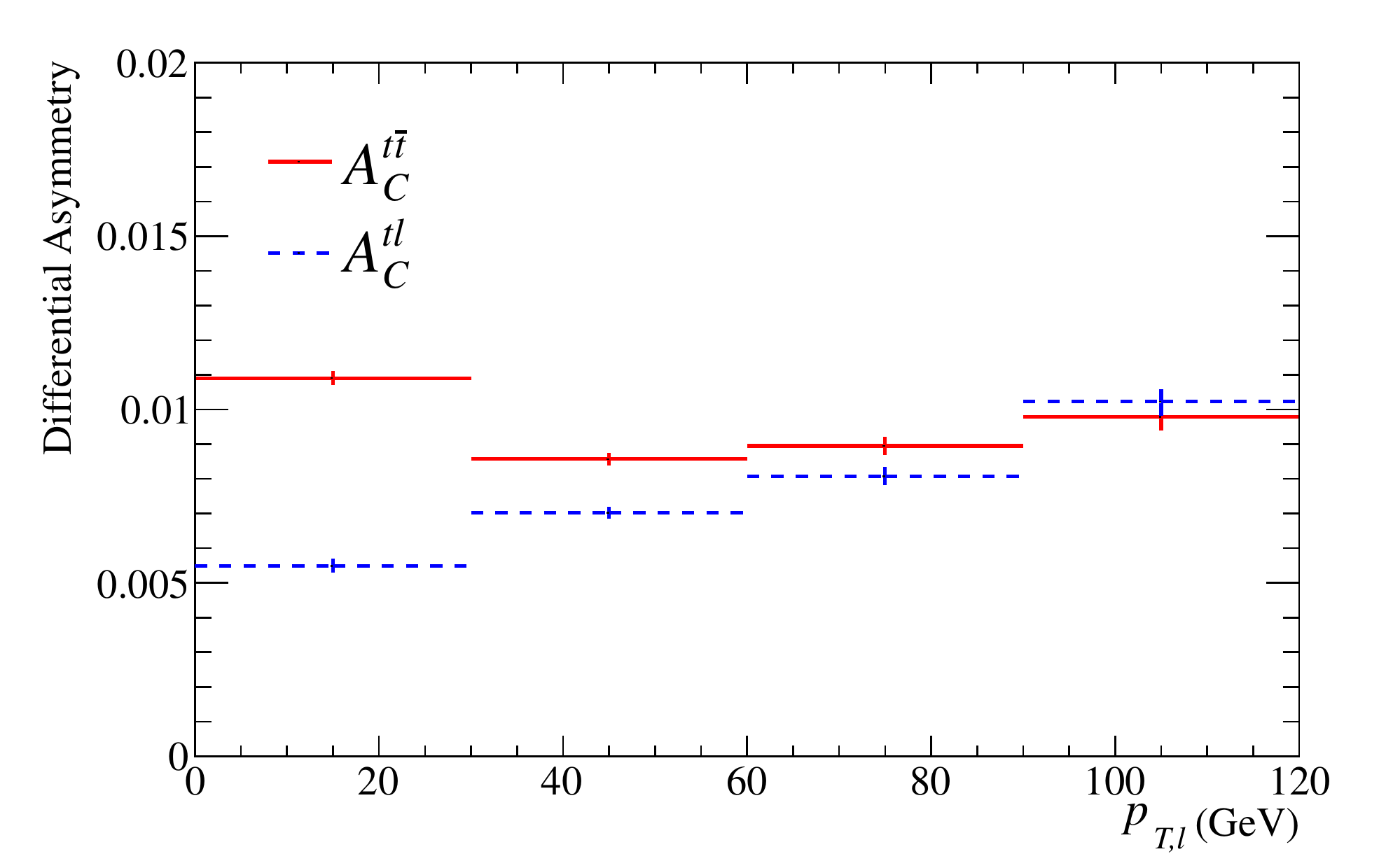}
\includegraphics[width=7cm,height=5cm]{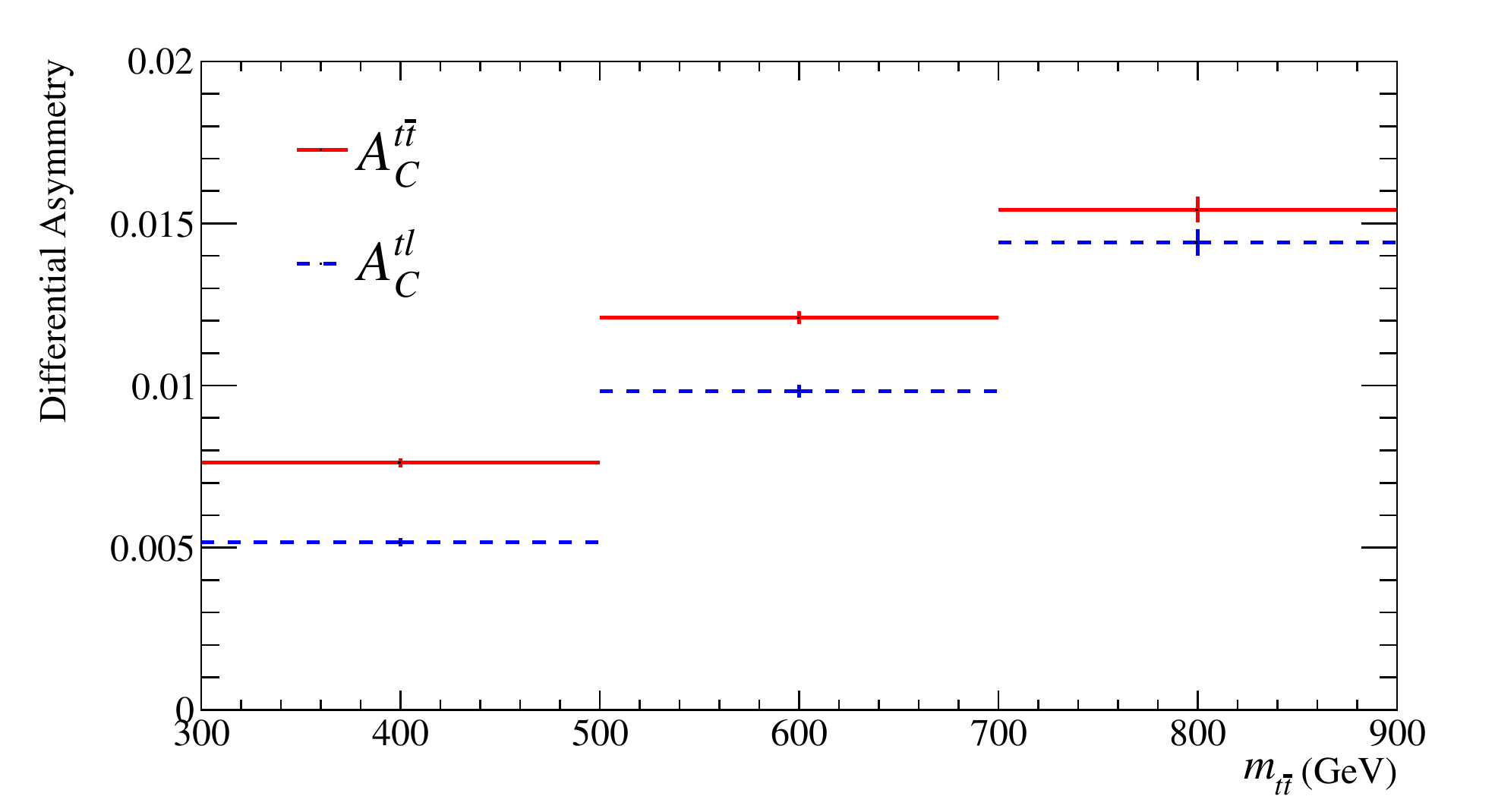}
\caption{Charge and lepton-based asymmetry dependence on $\ptl$ (left panel)
  and $\mttb$ (right panel) in the SM with no cuts applied. The error
  bars correspond to Monte Carlo statistical errors.
}\label{ac}
\end{figure}

As mentioned before, we will focus on the $\sqrt{s}=8$ TeV LHC run. 
We have generated our SM $\ttb$ events using the next-to-leading order (NLO) event generator POWHEG \cite{powheg}, with
the CT10 \cite{pdfct10} parton distribution functions and with the
renormalization and factorization scales set to $\mu_{R}=\mu_{F}=Q =
\sqrt{m_{t}^{2} +(p_{T,t})^{2}}$. 
The spin correlations between the top and anti-top quarks  and their
decay products are maintained in 
the simulated events. We show in Fig.~\ref{ac} 
the corresponding distributions for $\ACttb$ (red solid) and $\AClt$
(blue dashed) as a function of $\ptl$ and $\mttb$ in the left and
right panels, respectively, in the SM with no cuts applied.
As expected, $\AClt$ tends to $\ACttb$ at large $\ptl$ since leptons with a large transverse momentum come from the decay of boosted top quarks, which result in $y_{l}\approx y_{t}$. 
According to the right plot in Fig.\ref{ac},  both asymmetries $\ACttb$ and $\AClt$ grow with $m_{t\bar{t}}$. 
Since events with large lepton $p_{T}$ are  correlated  with large $m_{t\bar{t}}$,
 the lepton asymmetry $\AClt$ approaches  $\ACttb$  also at large invariant mass of $t\bar{t}$ pairs.

\begin{figure}[hbtp]
\centering
\includegraphics[width=7cm,height=5cm]{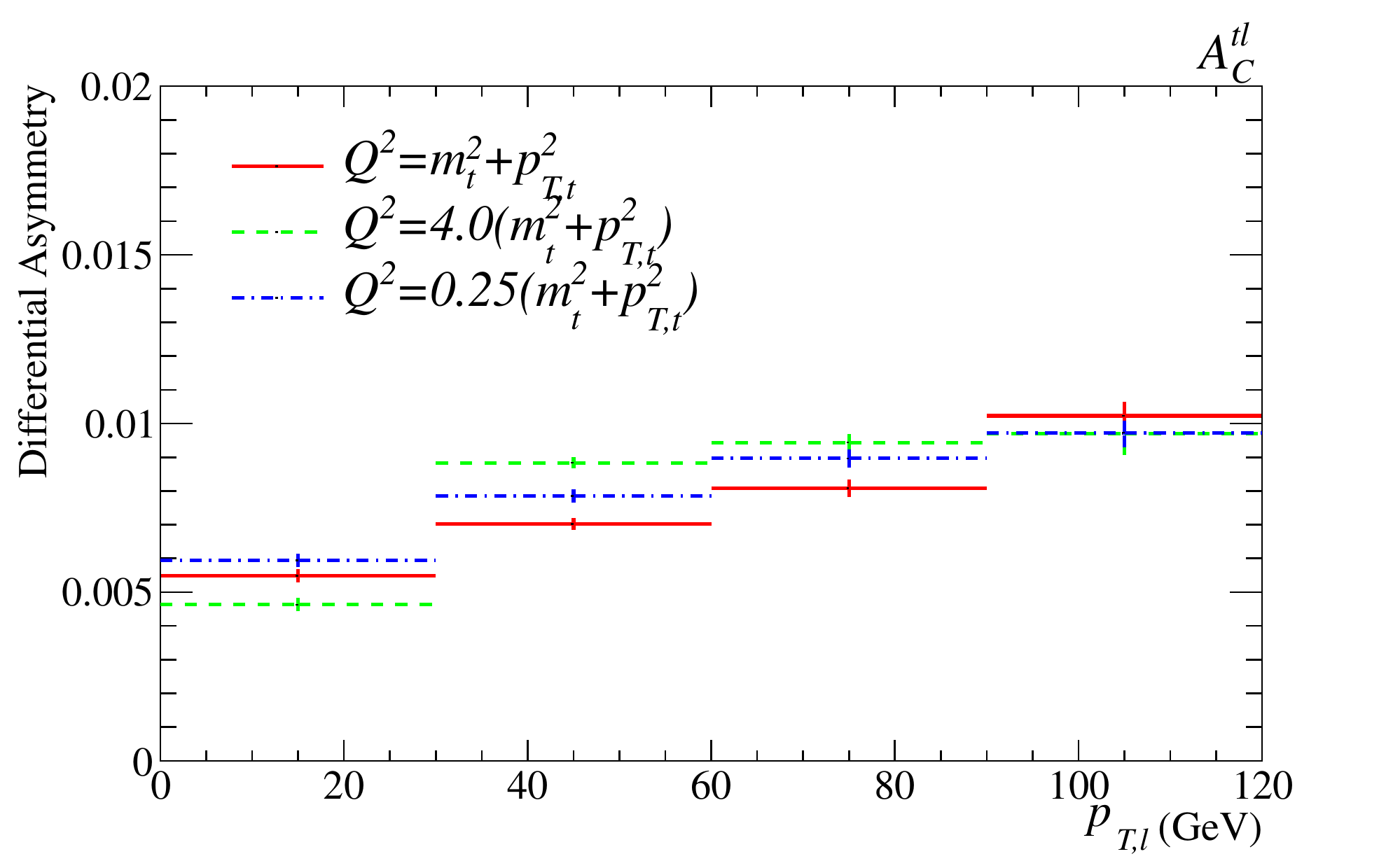}
\includegraphics[width=7cm,height=5cm]{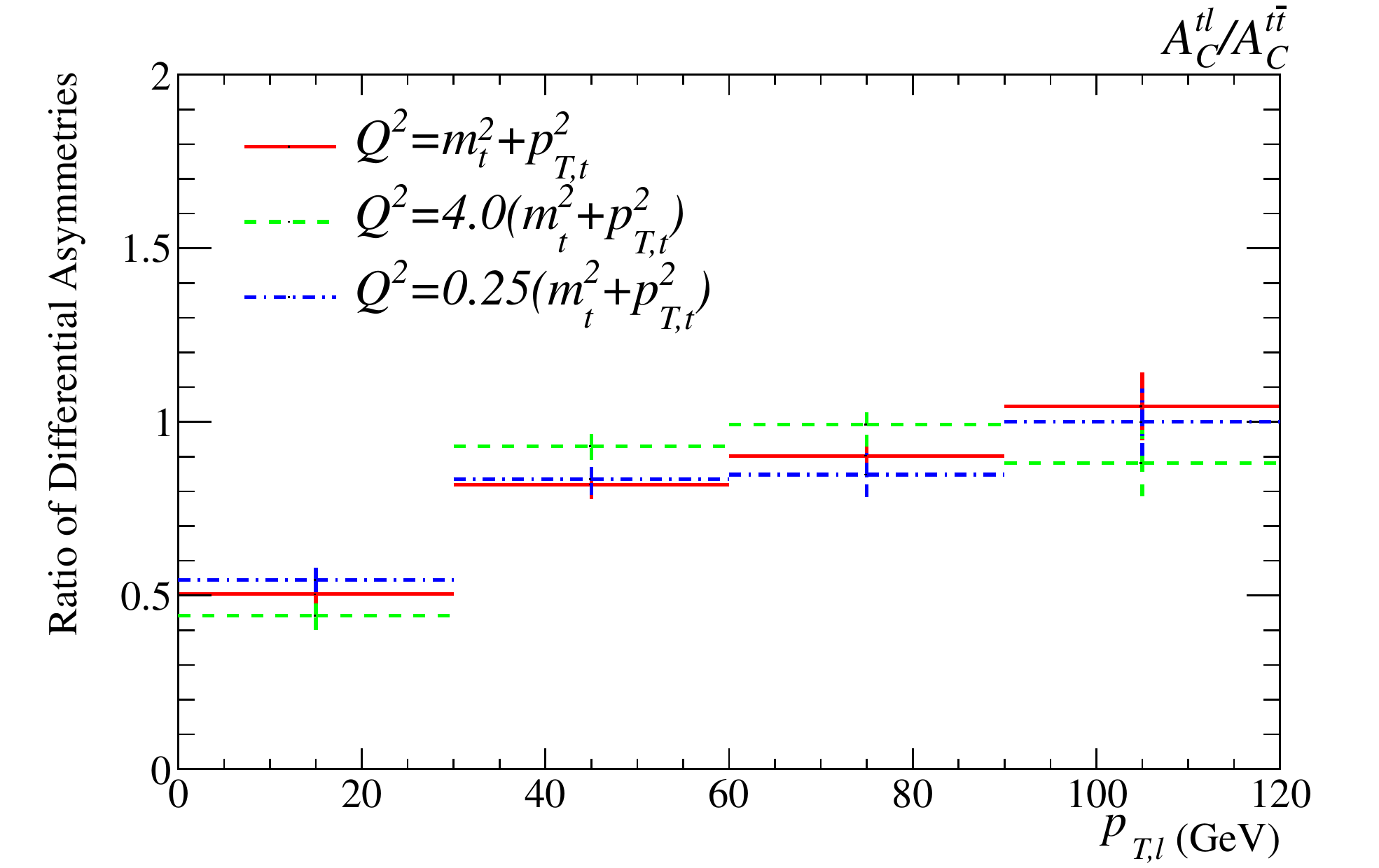}
\includegraphics[width=7cm,height=5cm]{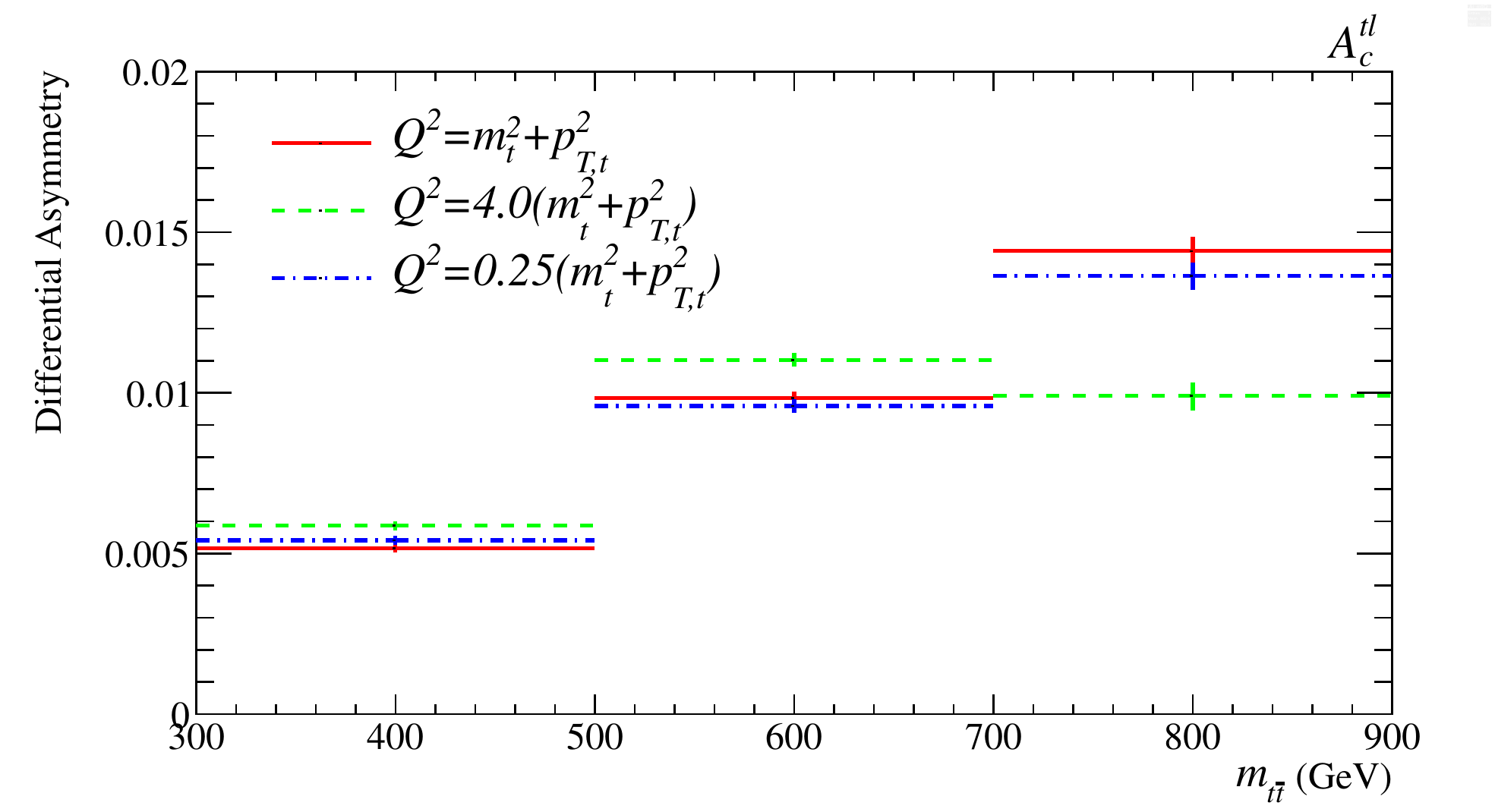}
\includegraphics[width=7cm,height=5cm]{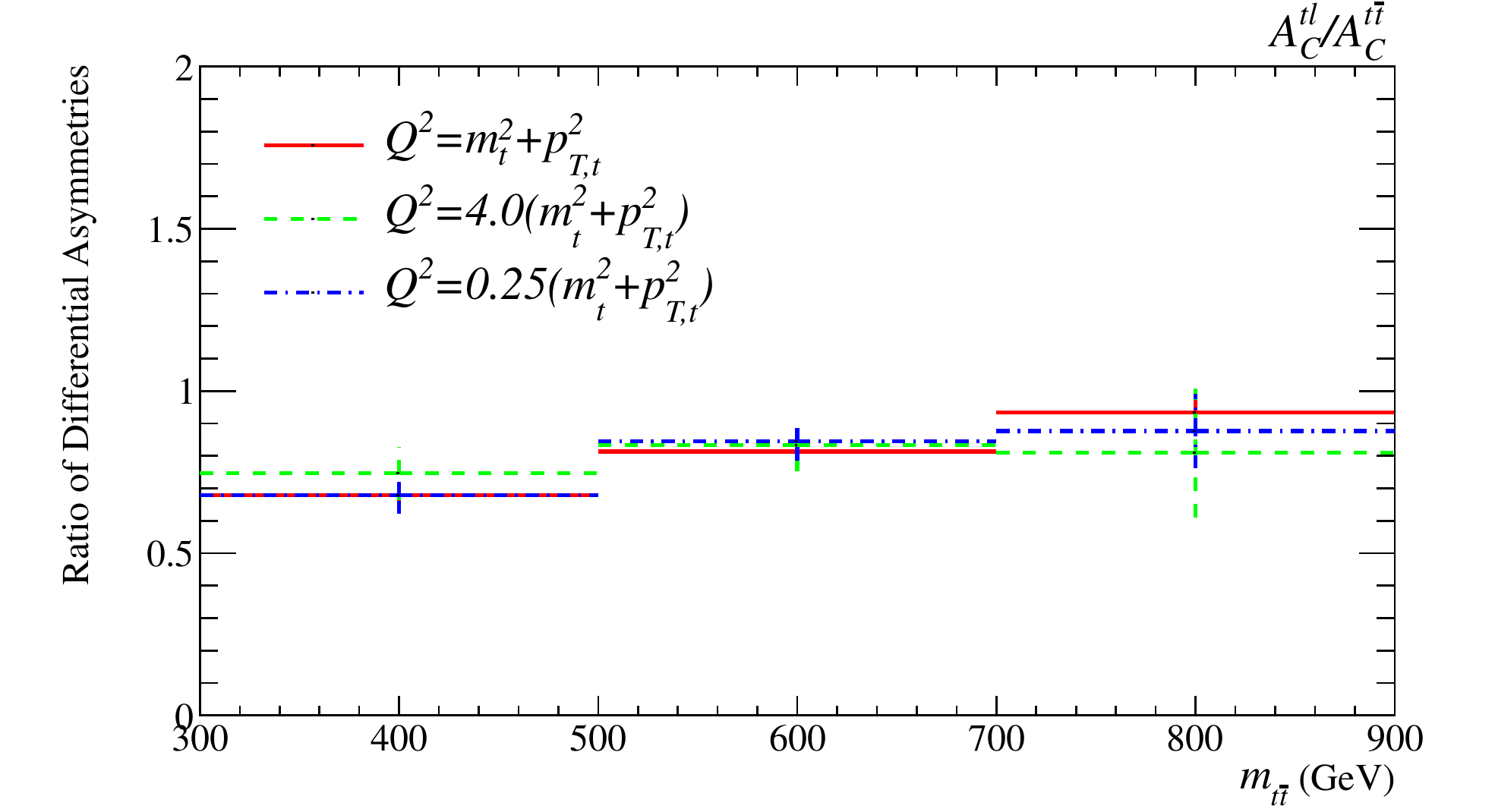}
\caption{Distribution of $\AClt$ (left) and the $\AClt/\ACttb$ ratio
  (right) as a function of $\ptl$ (top) and $\mttb$ (bottom) for three
  different choices of the renormalization and factorization scale, in
  the SM with no cuts applied. }\label{scale} 
\end{figure}

We now proceed to investigate the robustness of the ratio of these
asymmetries, measured as a function of the kinematical variables,
against various simulation and reconstruction effects.
As a first check we will test the dependence of the ratio on the
renormalization and factorization scales.
Since we are using a NLO calculation, which is the first order at
which the asymmetries are generated, 
we need to estimate the effects of ignoring higher-order corrections.
We have done that by increasing and reducing the scales in calculation
of the asymmetries by a factor of two. It is expected that
each asymmetry would show a sizable variation with the change of the scale
but  due to the correlation described above the asymmetries ratio would be stable under such variation.
We show in Fig.~\ref{scale}, the $\ptl$ (top) and $\mttb$ (bottom)
distributions of $\AClt$ (left) and of the $\AClt/\ACttb$ ratio (right)
for the three different choices of the renormalization and
factorization scales $Q^{2} = Q_{0}^{2}$, $Q^{2} = 4 \times Q_{0}^{2}$
and $Q^{2} = 0.25\times Q_{0}^{2}$ where $Q_{0}^{2} = m_{t}^{2}
+(p_{T,t})^{2}$. These results have been obtained in the SM with no
cuts applied. The two plots on the right of the figure show that the
ratio of asymmetries is indeed quite stable, consistent with the statistical Monte Carlo uncertainties, when measured as a function of
both $\ptl$ and $\mttb$.

\begin{figure}[hbtp]
\centering 
\includegraphics[width=7cm,height=5cm]{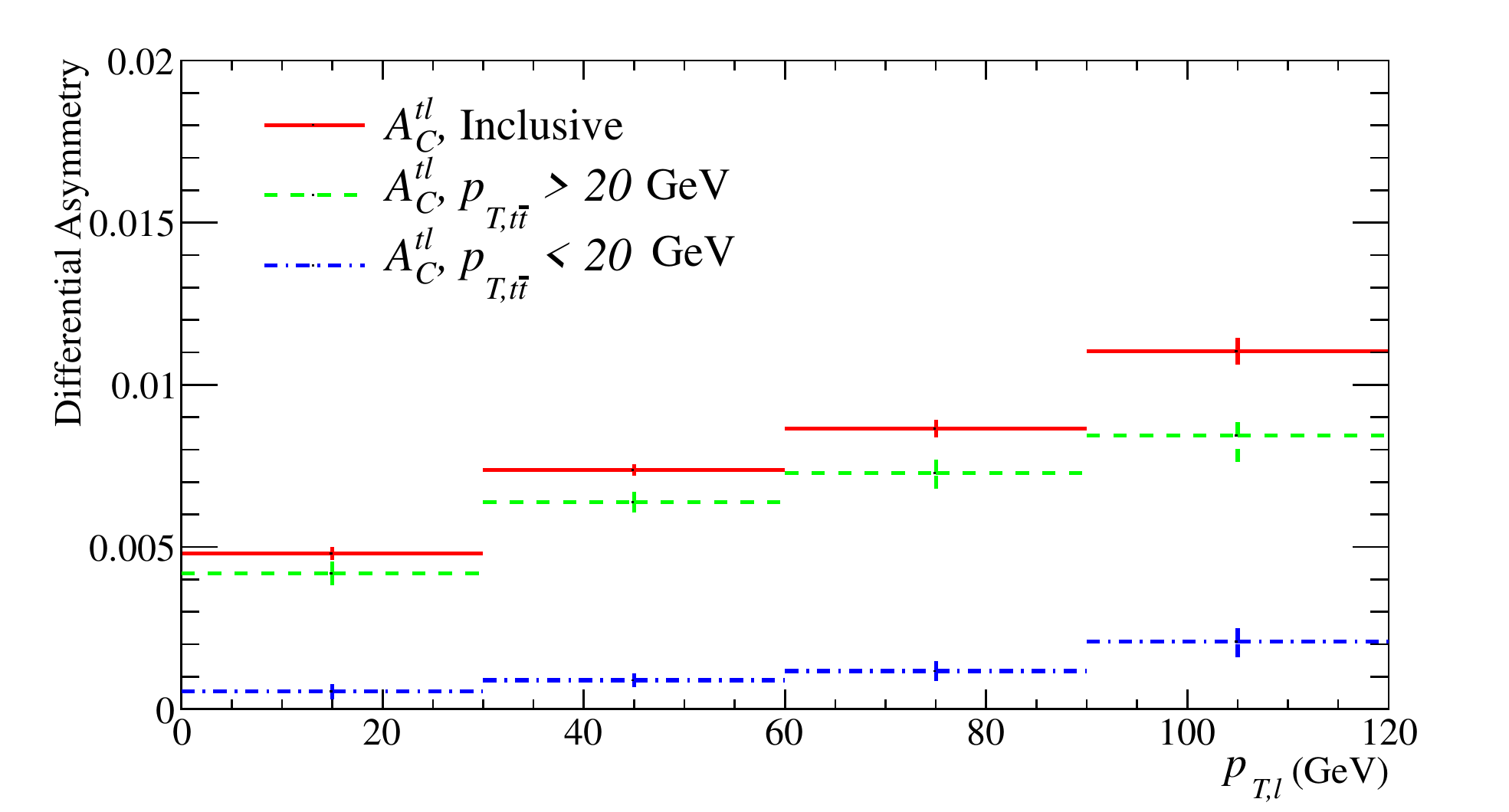}
\includegraphics[width=7cm,height=5cm]{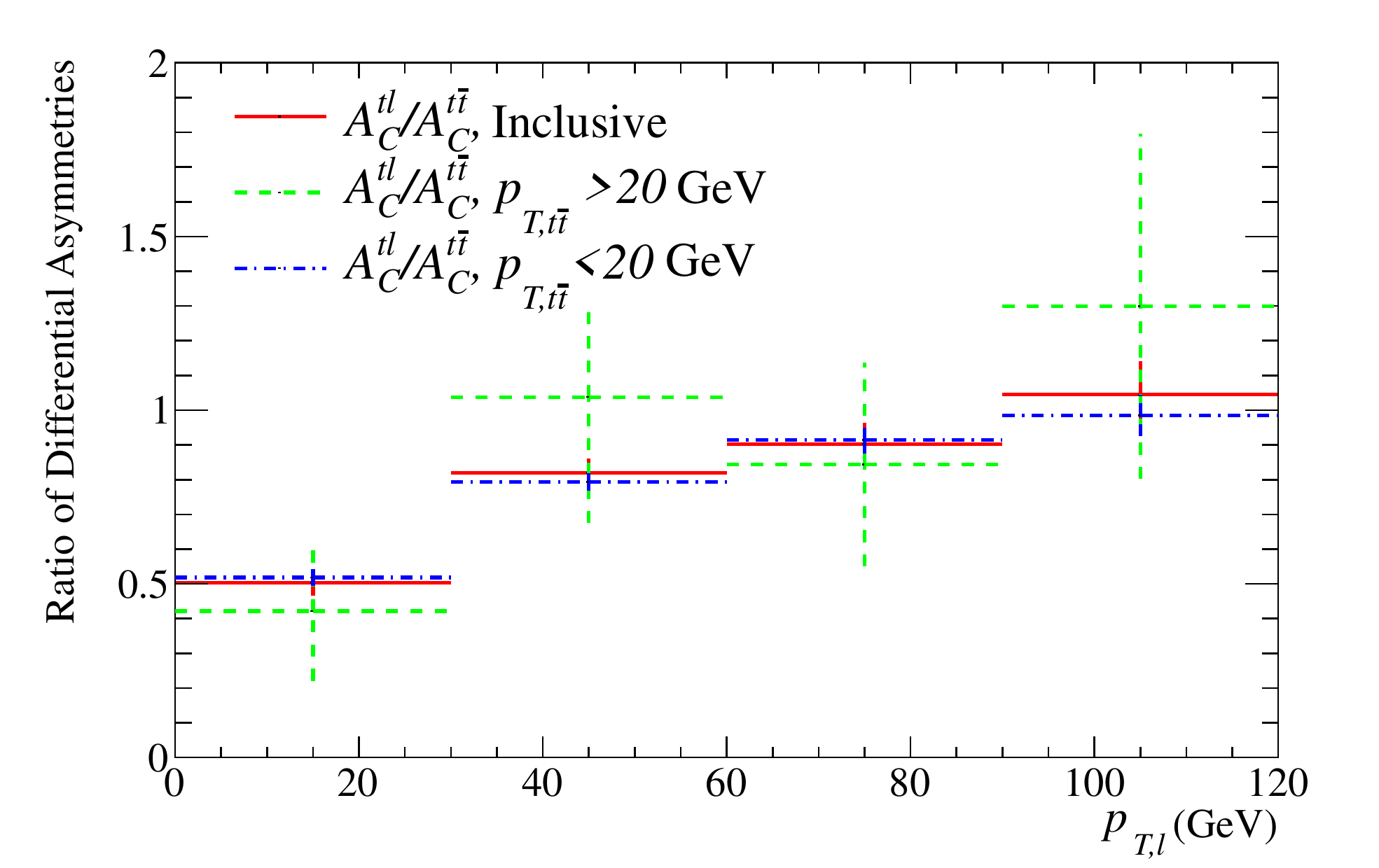}
\includegraphics[width=7cm,height=5cm]{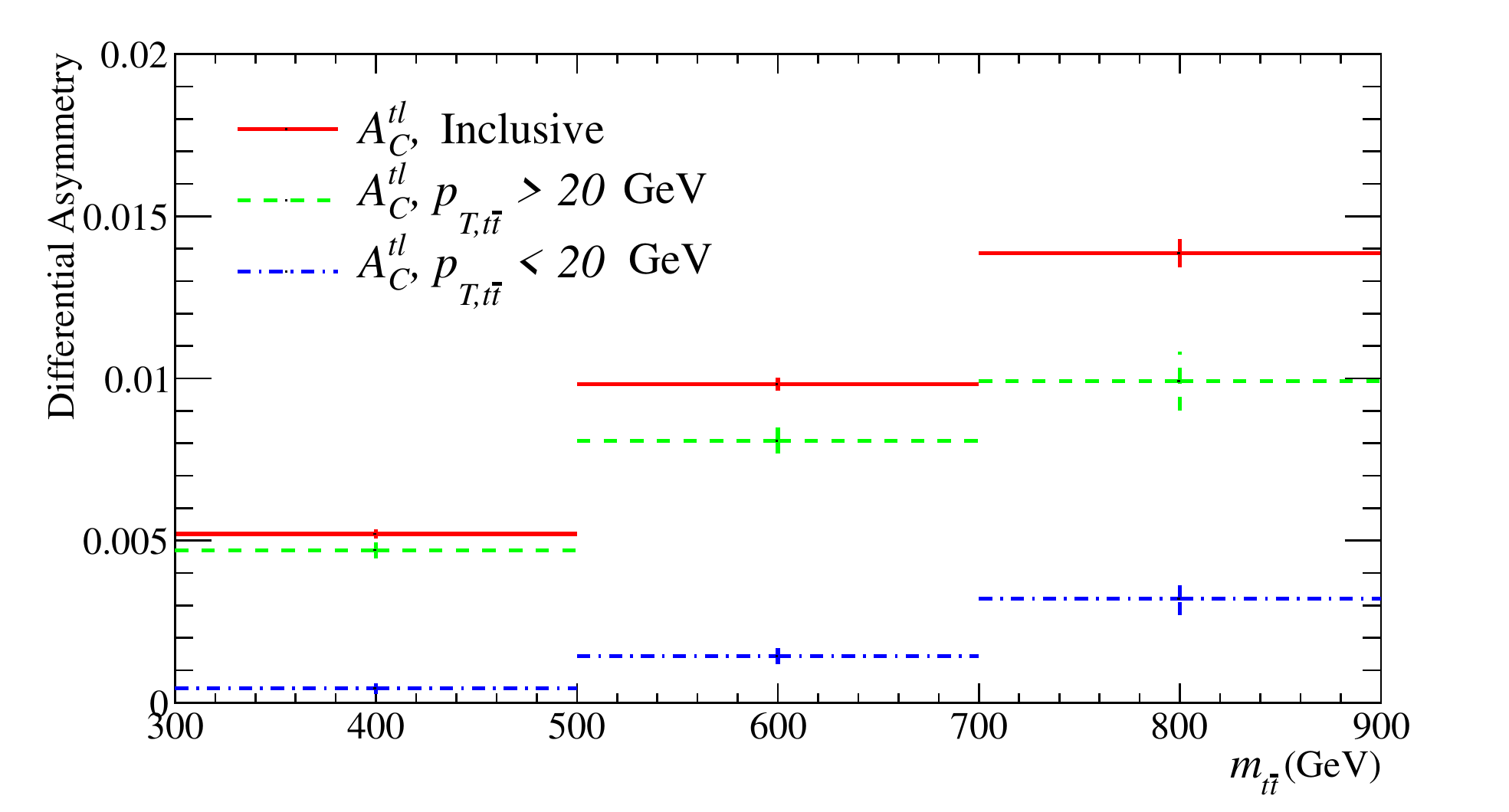}
\includegraphics[width=7cm,height=5cm]{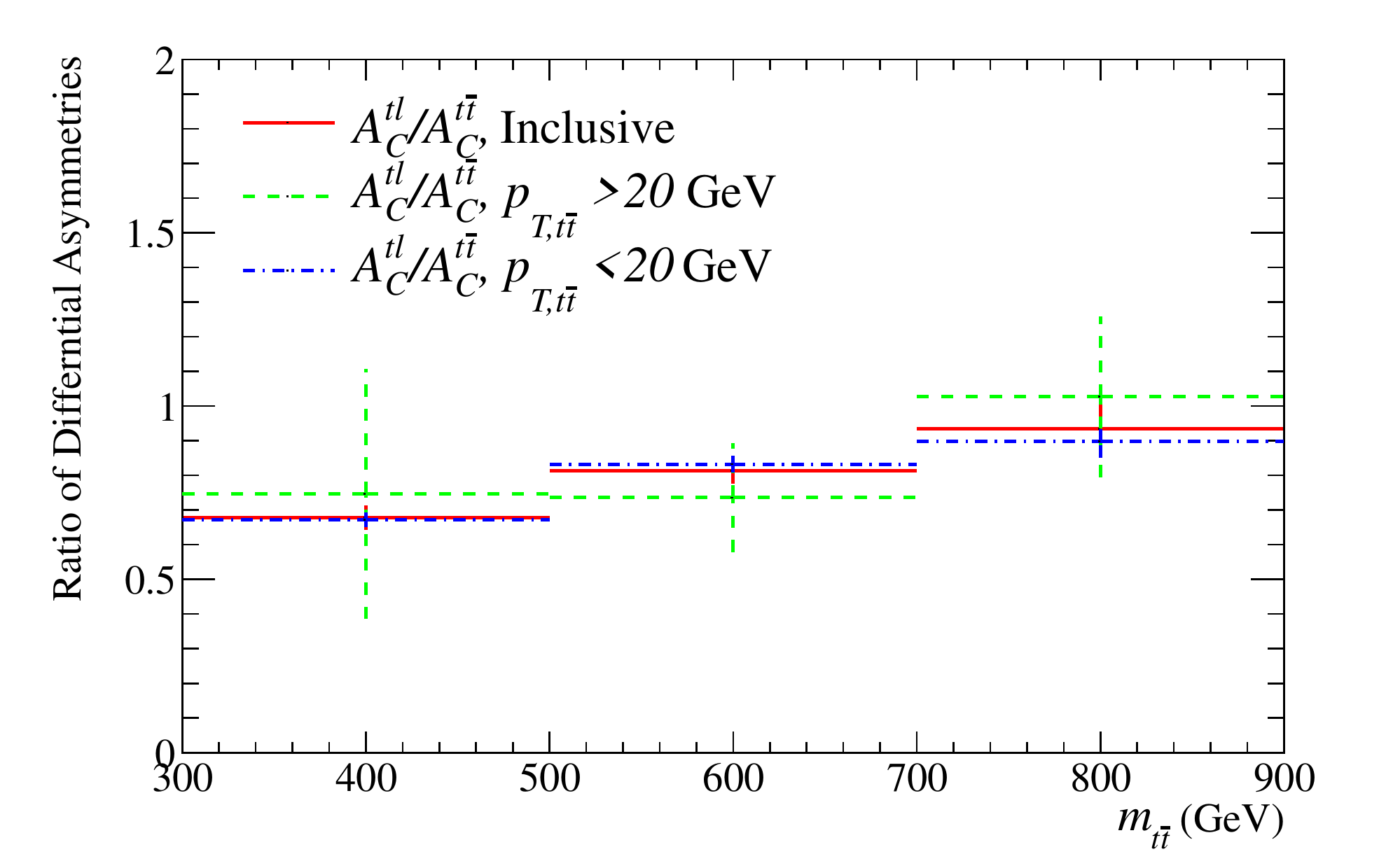}
\caption{Distribution of $\AClt$ (left) and the $\AClt/\ACttb$ ratio
  (right) as a function of $\ptl$ (top) and $\mttb$ (bottom)  for two different values of the $\ttb$ system
  transverse momentum: $p_{T,t\bar{t}}< 20$ GeV and $p_{T,t\bar{t}} > 20$ GeV. 
To calculate $\AClt$ for  $p_{T,t\bar{t}} > 20$ GeV we use the NLO cross section from POWHEG in the denominator because the LO one vanishes in this case. 
The largest part of the cross section is in the lower
 $p_{T,t\bar{t}}$ region, so the statistical uncertainties on the high
 $p_{T,t\bar{t}}$ region  are higher. }\label{ratiolptmtt}
\end{figure}

$\ACttb$ depends on the
transverse momentum of the $t\bar{t}$ system, $p_{T,t\bar{t}}$
(see for example~\cite{acpaper}). The reason is that $p_{T,t\bar{t}}$ is correlated with the amount of real emission in the events that together with the virtual corrections is inducing the asymmetries. Larger values
of $p_{T,\ttb}$ typically correspond to events with harder real radiation.
In the SM, the interference of the born and box diagrams in top pair
production contributes positively to $\ACttb$ while the interference
of diagrams with initial and final state radiation contributes
negatively. Thus, by varying the value of
$p_{T,t\bar{t}}$ and therefore the amount of hard real radiation, we
can modify the relative positive and negative contributions to the
asymmetry. Events with larger values of $p_{T,t\bar{t}}$ mostly
produce negative charge asymmetry. 
Thus, it is important to investigate whether 
the asymmetries are stable in two kinematic regimes with positive and
negative contributions to the charge asymmetry when measured as a
function of $\ptl$ and $\mttb$.
We show in Fig.~\ref{ratiolptmtt} $\AClt$ and the $\AClt/\ACttb$ ratio as a
function of $\ptl$ (left) and $\mttb$ (right) in two different
$p_{T,\ttb}$ regimes: $p_{T,t\bar{t}}<20$ GeV and $p_{T,t\bar{t}}>20$
GeV, together with the inclusive result. 
Again we see that the ratio is robust against the  value of the $p_T$ of the $\ttb$ system.

Before closing this part we would like to point  that we have performed an additional, new robustness test in this study in the context of the Tevatron measurements.
In the appendix, in subsection~\ref{Teva}, we investigate whether the SM correlation between 
the lepton-based asymmetry versus $p_{T}$ of lepton is sensitive to the use of the specific tool that is used to calculated the matrix elements and parton shower.
We compare the POWHEG~\cite{powheg}
and MC@NLO~\cite{mcnlo} event generators.
These event generators are suitable for this measurement since both include
the NLO calculation of top pair production with
subsequent simulation of parton showers.
The $t\bar{t}$ events generated with MC@NLO have been combined with
HERWIG~\cite{herwig}   for showering and hadronization 
and the POWHEG events have been combined with PYTHIA~\cite{pythia} for parton showering and
hadronization. As shown in Fig.~\ref{Ald0} a fantastic agreement between the two NLO tools is observed.

\section{Top versus Lepton Asymmetry beyond the SM} 
\label{sec:bsm}

As we have shown in the previous section, 
within the SM, the ratio $\AClt/\ACttb$ is rather insensitive to  theoretical uncertainties and reconstruction effects, 
and this robustness is true  for differential asymmetries as functions of $\mttb$ or $\ptl$. 
This is in contrast to  $\ACttb$ or $\AClt$ on their own,  where a
much larger variation of the predictions obtained  
with current Monte Carlo tools can be observed.       
As we show in this section, the ratio of differential asymmetries is
also a powerful discriminant between the SM and new physics models
explaining the Tevatron anomaly.
The reason is that, in the SM,  the lepton-based asymmetry is
inherited from the top asymmetry: the direction of the lepton in
semi-leptonic top decays is correlated with the direction of the
decaying top.  
Beyond the SM, however,  $\AClt$ becomes independent of $\ACttb$
because polarization effects in the $t \bar t$ production may affect
these two in a completely different way.  
This suggests we can use the {\em shape} of $\AClt/\ACttb$ as function
of $\mttb$ or $\ptl$ to differentiate between the SM and BSM interpretations of the measured asymmetries.  
In this section we illustrate this idea by calculating $\ACttb$ and $\AClt$  predicted by a set of BSM benchmark models.    
The criterion for choosing our benchmarks is that they should improve the global fit to the asymmetry observables and $t \bar t$ cross
section measurements  at the Tevatron and the LHC.   
In the following we first discuss the most relevant constraints, next
we introduce our benchmark models, and finally we list the results for
$\AClt$, $\ACttb$,  and their ratio in these models.

\subsection{Constraints}

New physics models contributing to the top asymmetry  are constrained
by measurements of the total and differential cross section at the
Tevatron and the LHC \cite{AguilarSaavedra:2011hz,AguilarSaavedra:2011ug,Fajfer:2012si}.  To design  our benchmarks we have taken into
account the following constraints:   
\begin{enumerate} 
\item The Tevatron combination of the  $t \bar t$ inclusive cross section \cite{Aaltonen:2013wca}: 
\beq
\label{eq:stt}
\sigma_{t\bar t}^{\rm TeV} = (7.62 \pm 0.42)~\textrm{pb},   
\eeq  
where  the SM next-to-next-to-leading order (NNLO) prediction is  $\sigma_{t\bar t, \ \rm SM}^{\rm TeV} = 7.16^{+0.20}_{-0.23}~\textrm{pb}$ \cite{Czakon:2013goa}. 
\item The last bin of the CDF \cite{Aaltonen:2009iz} and D$\O$  \cite{D0mtt} differential $t \bar t$ cross section measurement as a function of $m_{t \bar t}$:   
\bea 
{\bf CDF:} \qquad \int_{0.8 \textrm{TeV}}^{1.4 \textrm{TeV}} {d \sigma_{t\bar t}^{\rm TeV} \over d m_{t \bar t}}  &=&  (0.041 \pm 0.21)~\textrm{pb},  
\nn 
{\bf D\O:} \qquad \int_{0.75 \textrm{TeV}}^{1.2 \textrm{TeV}} {d \sigma_{t\bar t}^{\rm TeV} \over d m_{t \bar t}}  &=&  0.067^{+0.052}_{-0.050}~\textrm{pb},  
\eea 
where the SM prediction is quoted as $\int_{0.8 \textrm{TeV}}^{1.4 \textrm{TeV}} {d \sigma_{t\bar t, \rm SM}^{\rm TeV} \over d m_{t \bar t}} \approx 0.03~\textrm{pb}$, and $\int_{0.75 \textrm{TeV}}^{1.2 \textrm{TeV}} {d \sigma_{t\bar t, \rm SM}^{\rm TeV} \over d m_{t \bar t}} \approx 0.06~\textrm{pb}$.  
\item The  95\% CL limit on  the $t \bar t$ cross section at the high $\mttb$ tail at CMS \cite{Chatrchyan:2013lca}: 
\beq
\label{eq:cmstail}
{ \int_{1 \textrm{TeV}}^{8 \textrm{TeV}} \Delta {d \sigma_{t\bar t}^\textrm{TeV} \over d m_{t \bar t}}  \over  \int_{1 \textrm{TeV} }^{8 \textrm{TeV} } \Delta {d \sigma_{t\bar t,\rm SM}^\textrm{TeV}  \over d m_{t \bar t}}  } < 1.2~. 
\eeq 
\end{enumerate}

\subsection{Benchmark models}

One class of BSM models generating the top forward--backward asymmetry at tree-level contains a color-octet vector boson $G_\mu^a$  (the so-called {\em axigluon}) with non-zero mass $m_G$ and chiral couplings \cite{Hall:1985wz}. 
The axigluon couplings to the SM quarks are assumed to be flavor diagonal but otherwise arbitrary:
\begin{equation}
\label{eq:axigluon}
{\cal L}\;\supset\;g_{L,i}\,\bar q_i\,\gamma^\mu\,G_\mu^a\,T^a P_L \,q_i\;+\;
g_{R,i}\,\bar q_i\,\gamma^\mu\,G_\mu^a\,T^a\, P_R \,q_i\,,
\end{equation}
where $q_i$ are the SM quarks fields, and $P_{L,R}$ are the projection operators into left- and right-handed spinors. 
In this model the top pair production amplitude $q \bar q \to t \bar t$ receives a contribution from the axigluon in the s-channel which interferes with the SM gluon exchange.
The forward-backward asymmetry appears at tree level when the axigluon couplings are chiral.     
We choose several benchmarks with different axigluon mass and couplings. 
First, we choose three benchmarks with a light axigluon:  
\bea
{\rm \bf Axi200R:}  &\quad & m_G=200~\textrm{GeV}, \quad \Gamma_G=50~\textrm{GeV},
\quad  g_{R,i}= 0.5 g_s, \quad g_{L,i} = 0;   
\nn 
{\rm \bf Axi200L:}  &\quad & m_G=200~\textrm{GeV}, \quad \Gamma_G=50~\textrm{GeV},
\quad  g_{R,i}= 0,\phantom{.4g_s} \quad g_{L,i} = 0.5 g_s; 
\nn 
{\rm \bf Axi200A:}  &\quad & m_G=200~\textrm{GeV},   \quad \Gamma_G=50~\textrm{GeV},
\quad  g_{R,i}= 0.4 g_s, \quad g_{L,i} = -0.4 g_s,  
\eea  
where $g_s$ is the strong coupling. 
A light axigluon, $100$~GeV $\lesssim m_G \lesssim 400$~GeV gives rise
to a  positive asymmetry when couplings are flavor universal as in
\cite{Tavares:2011zg,AguilarSaavedra:2011ci}. Such a particle can be
consistent with all existing constraints as long as it has a
significant width \cite{AguilarSaavedra:2011ck,Gross:2012bz,Gresham:2012kv}. In the benchmarks above we set
$\Gamma_G=50$~GeV, even though the decay width into the SM  is only
${\cal O}$(few)~GeV; the remaining width must come from exotic
(e.g. multijet) axigluon decays channels \cite{Gross:2012bz}.   
Compared to the similar benchmarks studied in  \cite{Falkowski:2012cu},  Axi200R and Axi200L  have reduced couplings in order to reduce the tension with the total Tevatron cross section and lepton-based asymmetry measurements, at the price of a smaller contribution to the $t \bar t$ asymmetry. 

We also choose 2 benchmarks with a heavy axigluon:
\bea 
{\rm \bf Axi2000A:}  & \ & m_G=2~\textrm{TeV},  \quad \Gamma_G=0.96~\textrm{TeV}, \quad 
g_{R,u}= - g_{L,q_1} =  -0.6 g_s,   \quad g_{R,t} =  - g_{L,t} = 4 g_s; 
\nn 
{\rm \bf Axi2000R:}  &\ & m_G=2~\textrm{TeV},  \quad \Gamma_G=1.0~\textrm{TeV}, \quad  
g_{R,u}=  -0.8 g_s,   \quad g_{R,t} = 6 g_s, \quad g_{L,i}=0. 
\eea 
For a heavy axigluon obtaining a  positive asymmetry requires flavor
non-universal couplings, in particular the sign of the coupling to the
light and top quarks has to be opposite. In this case $\Gamma$ is
equal to the decay width into the SM quarks.  
The mass of about 2 TeV is needed to avoid the constraints from the $t
\bar t$ at the LHC, unless new decay channels provide a large
width~\cite{Barcelo:2011vk}.  The couplings to light quarks must be moderate to
avoid dijet bounds, but then to achieve a
significant contribution to 
the top asymmetry the coupling to the  top quark must be close to  the
non-perturbative regime.  

Finally, we consider a different model with  a complex gauge boson
$Z'_\mu$ coupled to right-handed up-type quarks in a flavor-violating
way,  
\beq
{\cal L}\;\supset\; g_{Z'} Z_\mu' \bar t_R \gamma^\mu u_R + \hc .   
\eeq 
$Z'$ needs to be complex \cite{Jung:2011zv}, otherwise generating a large top asymmetry is not possible without conflicting the bounds from the  same-sign top  production   \cite{Jung:2009jz,AguilarSaavedra:2011zy}. 
The new gauge boson contributes to the  $u \bar u \to t \bar t$  in
the t-channel which yields positive contribution to the top asymmetry
if $g_{Z'}$ is large enough (for a small $g_{Z'}$ the contribution is
negative). Furthermore, it also contributes to  $g u/\bar u \to t \bar
t  u$ process via an on-shell  $Z'$ production followed by the decay $Z'
\to t \bar u$ and its conjugate~\cite{Drobnak:2012rb}. The latter
process  is negligible at the Tevatron, but becomes important at the
LHC where the available phase space and the gluon luminosity are
larger.   
We choose the benchmark point as 
\bea &
{\rm \bf Zp220:} \qquad m_{Z'}=220~\textrm{GeV},   \qquad g_{Z'}=  0.7,
\qquad \Gamma_G=2.9~\textrm{GeV}.  
\eea 
The mass and the coupling are chosen such that a sizable Tevatron top
asymmetry is generated. However at the LHC the asymmetry approximately
cancels between $u \bar u \to t \bar t$ (contributing with a positive
sign) and $g u \to t  Z' \to t \bar t u$  (contributing with a
negative sign).     

In Table~\ref{tab:benchmarks} we collect the additional contribution
to the inclusive asymmetries at the
Tevatron and the LHC predicted for all the  benchmarks introduced
above.  
 
\begin{table} 
\begin{center}
\begin{tabular}{l|c|c|c|c|}
Benchmark & $\Delta \Attb$ &  $\Delta \Al$ &  $\Delta \ACttb$ & $\Delta \AClt$    
 \\ \hline 
{\bf Axi200R }& 0.05 &   0.07 & 0.006 & 0.009 
 \\  \hline  
{\bf Axi200L }& 0.05 &   -0.03  & 0.007 & 0.001 
\\ \hline 
{\bf  Axi200A }& 0.12 & 0.05 & 0.016 & 0.012 
\\  \hline 
{\bf  Axi2000R }&  0.04 & 0.05 & 0.007 & 0.009
\\  \hline 
{\bf  Axi2000A }& 0.07 & 0.04 & 0.012 & 0.010
\\ \hline 
{\bf Zp220} & 0.13 & 0.02 & -0.001 & 0.005  
\end{tabular}
\caption{Additional contribution to inclusive top and lepton-based asymmetries at the Tevatron and the LHC for the benchmarks studied in this paper.}
\label{tab:benchmarks}
\end{center}
\end{table}

\subsection{Results}

In Tables~\ref{tab:actbsm}~and~\ref{tab:acltbsm} we give our results
for the charge and lepton asymmetries at the 8 TeV LHC for different
$\ptl$ and $\mttb$ bins in the 6 BSM benchmarks considered.
We have obtained these numbers in the following way. We have computed
the leading order (LO) 
BSM correction to the forward and backward cross sections  in
each bin using MadGraph~5 \cite{Alwall:2011uj}.     
These were added to the NLO SM  
forward and backward cross sections computed with POWHEG. 
Finally, the asymmetry was obtained by taking the ratio of the difference of
the forward and backward cross sections divided by the sum of the 
LO cross sections in each bin.   

\begin{figure}[hbtp]
\centering
\includegraphics[width=0.45\textwidth]{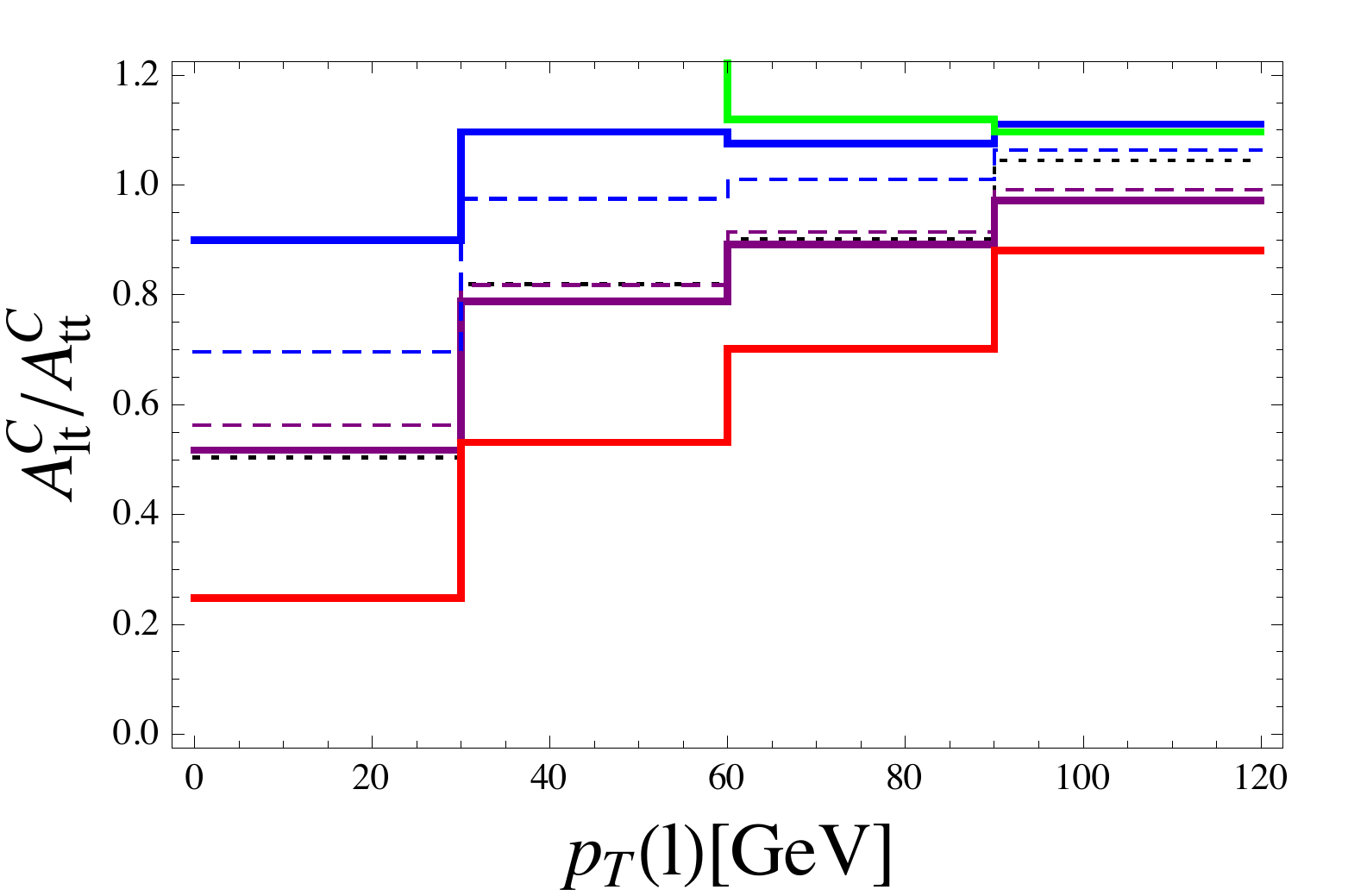}
\includegraphics[width=0.45\textwidth]{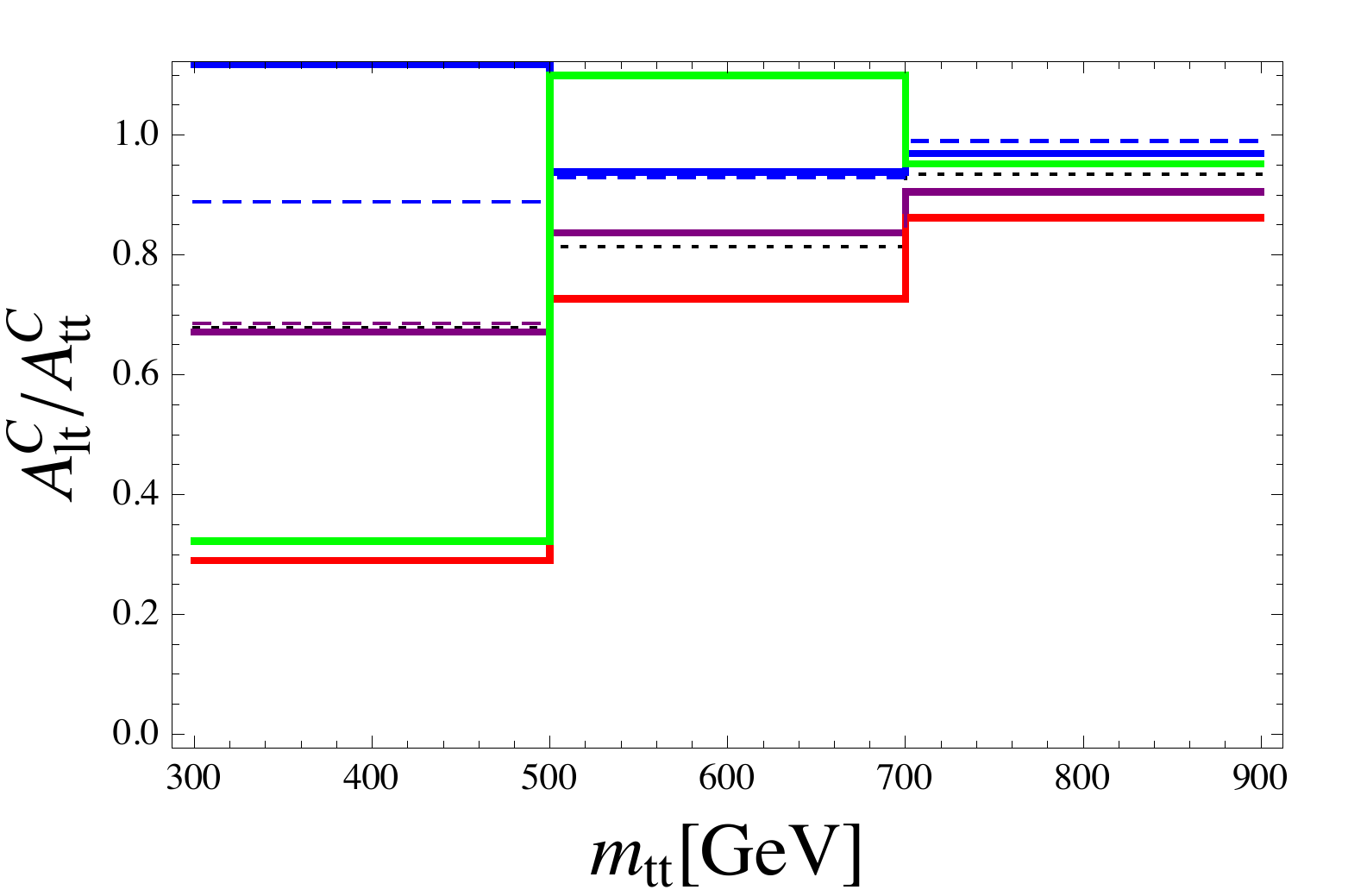}
\caption{Distribution of the ratio $ \AClt/\ACttb$ at the LHC as a
  function of $\ptl$ (left) and $\mttb$ (right) for the SM (dotted
  black)  and for the BSM benchmarks studied in this paper: 
Axi200R (solid blue), Axi200L (solid red), Axi200A (solid purple), Axi2000R (dashed blue), Axi2000A (dashed purple), and Zp220 (solid green).     
 }
 \label{fig:rbsm}
\end{figure}

As we have stressed previously, the most interesting observable is the
$\AClt/\ACttb$ ratio, that we show in Fig~\ref{fig:rbsm} as a function
of $\ptl$ (left) and $\mttb$ (right).   
We can see that the discriminating power of this observable,
previously pointed out in the context of the Tevatron asymmetry
\cite{Falkowski:2012cu}, survives at the LHC.  
For the light axigluon benchmarks Axi200L and Axi200R the shape of the
$\AClt/\ACttb$ curve is completely different than in the SM.  
This is because for these benchmarks  $\AClt$ and $\ACttb$ are less
correlated with each other, especially in low $\ptl$ and $\mttb$ bins
where polarization  
 effects dominate over purely kinematic effects.  
 A similar  albeit weaker effect can be observed for  the heavy
 axigluon benchmark Axi2000R.  
 The new physics corrections to  $\AClt/\ACttb$  are even more
 dramatic for the $Z'$ benchmark Zp220 because, in addition,   $\ACttb$
 is affected by an accidental cancellation between off-shell and
 on-shell $Z'$  amplitudes. 
As a consequence,  the ratio of the asymmetries  in the two lowest $\ptl$ bins is very large (out of the plot in Fig.~\ref{fig:rbsm}).  
In this case the precise value of the ratio is not relevant, since it is very sensitive to changing the parameters of the model and also to Monte Carlo uncertainties. 
 However the large magnitude is an observable effect of the accidental cancellations in $\ACttb$ (without corresponding cancellations in $\AClt$ ) which  could be the smoking gun of new physics.    
 On the other hand, for the axigluon benchmarks with axial couplings
 (where there is no overall polarization in the initial or final state)
 the shape of $\AClt/\ACttb$ closely resembles that in the SM.  Hence
 in these 2 particular  cases the ratio is not a good discriminant
 between SM and BSM interpretations  of the $\Attb$ anomaly.   

\begin{table} 
\begin{tabular}{l|c|c|c|c|}
$\ptl$[GeV] & [0,30]  &  [30,60] &  [60,90] & [90,120]    
 \\ \hline 
{\bf Axi200R }&  0.015 & 0.014 & 0.017 & 0.019
 \\  \hline  
{\bf Axi200L }&  0.020 & 0.016  & 0.015 & 0.015 
\\ \hline 
{\bf  Axi200A }& 0.026 & 0.024 &  0.025 & 0.028
\\  \hline 
{\bf  Axi2000R }&  0.014 & 0.012 & 0.016 & 0.023 
\\  \hline 
{\bf  Axi2000A }& 0.021 & 0.018 & 0.021 & 0.027 
\\ \hline 
{\bf Zp220 } & 0.000 & 0.004 & 0.014 & 0.024 
\end{tabular}
\quad 
\begin{tabular}{l|c|c|c|c|}
$\mttb$[GeV]  & [300,500] &  [500,700] &  [700,900]   
 \\ \hline 
{\bf Axi200R }& 0.013 & 0.019 & 0.024 
 \\  \hline  
{\bf Axi200L }&  0.014 & 0.020  & 0.025 
\\ \hline 
{\bf  Axi200A }& 0.021 & 0.030 &  0.038 
\\  \hline 
{\bf  Axi2000R }& 0.010 & 0.019 & 0.035 
\\  \hline 
{\bf  Axi2000A }& 0.013 & 0.027 & 0.052 
\\ \hline 
{\bf Zp220 } & -0.011 & 0.008 & 0.069 
\end{tabular}
\caption{$\ACttb$ as a function of $\ptl$ (left) and $\mttb$ (right) for the benchmarks studied in this paper. }
\label{tab:actbsm}
\end{table}

\begin{table} 
\begin{tabular}{l|c|c|c|c|}
$\ptl$[GeV]  & [0,30] &  [30,60] &  [60,90] & [90,120]    
 \\ \hline 
{\bf Axi200R }& 0.013 & 0.015 & 0.018 & 0.021 
 \\  \hline  
{\bf Axi200L }& 0.005 & 0.008 & 0.011 & 0.013 
\\ \hline 
{\bf  Axi200A }& 0.013 & 0.019 &  0.023 & 0.027
\\  \hline 
{\bf  Axi2000R }&  0.010 & 0.012 & 0.016 & 0.024 
\\  \hline 
{\bf  Axi2000A }& 0.012 & 0.015 & 0.019 & 0.026 
\\ \hline 
{\bf Zp220 } & 0.007 & 0.009 & 0.015 & 0.027 
\end{tabular}
\quad
\begin{tabular}{l|c|c|c|c|}
$\mttb$[GeV]  & [300,500] &  [500,700] &  [700,900]   
 \\ \hline 
{\bf Axi200R }& 0.014 & 0.018 & 0.024 
 \\  \hline  
{\bf Axi200L }& 0.004 & 0.014 & 0.022 
\\ \hline 
{\bf  Axi200A }& 0.014 & 0.026 &  0.035 
\\  \hline 
{\bf  Axi2000R }&  0.009 & 0.018 & 0.035 
\\  \hline 
{\bf  Axi2000A }& 0.009 & 0.023 & 0.047 
\\ \hline 
{\bf Zp220 } & -0.004 & 0.008 & 0.065
\end{tabular}

\caption{$\AClt$ as a function of $\ptl$ (left) and $\mttb$ (right)
  for the benchmarks studied in this paper.} 
\label{tab:acltbsm}
\end{table}

\section{Conclusions} \label{sec:conclusions}

Tevatron measurements of the forward-backward asymmetry in $t\bar{t}$ production and related lepton-based asymmetries show an intriguing excess over the SM prediction.  In order to discriminate between the SM and new physics explanations of the anomaly it is desirable to employ observables that are robust with respect to theoretical uncertainties and reconstruction effects. 
It was recently argued~\cite{Falkowski:2012cu} that one observable with these properties is the differential ratio of the forward-backward lepton-based  and $t \bar t$ asymmetries at the Tevatron.  
In this article we defined a new lepton-based asymmetry at the LHC and showed that the ratio of this asymmetry and the $\ttb$ charge asymmetry, measured as a function of the $p_T$ of the lepton in semi-leptonic channel or the $\ttb$ pair invariant mass,   fulfills all the requirements of a robust observable.  
In particular, we have shown that the ratio depends weakly on the
renormalization and factorization scales  (that is to say, it is
expected to be stable against higher-order QCD corrections), and on
the amount of hard radiation in the process (measured by the $p_T$ of the $\ttb$ system). We also compared the differential ratio obtained by POWHEG and MC@NLO. The two NLO tools are in fantastic agreement regarding the predicted value for this ratio of asymmetries.

Furthermore, the ratio of lepton-based and $t \bar t$ charge asymmetries can be a powerful probe of new physics.  
We have considered a number of benchmark models beyond the SM  that improve the agreement with current experimental data. 
The benchmarks studied in this paper include light and heavy axigluon models with different coupling structure that
contribute to the asymmetry in the $s$-channel and a model with a
complex $Z^\prime$ gauge boson that provides a contribution in the $t$-channel from associate production processes. 
We have shown that, in the cases in which the chiral structure of the new physics process is different from the one in the SM (which is unpolarized) the ratio of the asymmetries shows a dependence on the kinematic variables strikingly different from the one in the SM. 
In the case of the $Z^\prime$ benchmark  an accidental cancellation between two different contributions to the $t \bar t$ charge asymmetry  makes the differences even more remarkable. 

Our studies have been performed for the LHC with $\sqrt{s}=8$ TeV center-of-mass energy. 
Nevertheless, the shape of the ratio of the asymmetries as function of $\ptl$ or $\mttb$ should be a particularly useful
observable for the longer LHC run with an upgraded energy $\sqrt{s}=13/14$ TeV. 
Moreover, we expect that the ratio of related asymmetries in the di-leptonic $t \bar t$ channel has similar robustness properties and discriminating power.

\section*{Acknowledgements}
We thank Adam Martin for discussions and help with the Monte Carlo simulations. 
M.C. and S.K.  would like to thank CERN Theory Group for hospitality
during the completion of this project. 
S.K. and M.M.N. would like to thank of  J. Wagner-Kuhr and T. Chwalek
for help with the Monte Carlo simulations.  
J.S. would like to thank M.Perez-Victoria for useful comments on the
manuscript.  
A.C. is supported by the Swiss National Science Foundation under
contract SNF 200021-143781. 
M.C. and J.S. are supported by  MINECO grants AIC-D-2011-0690,
FPA2006-05294, FPA2010-17915 and the FPU program (M.C.) and by Junta de
Andaluc\'{\i}a grants FQM 101 and FQM 6552. 
The work of A.F. was supported by the ERC advanced grant Higgs@LHC.  
G.P. is supported by GIF, Minerva, IRG, ISF grants and by the Gruber award.

\appendix

\section{Forward-backward and lepton-based asymmetries at the
  Tevatron}

The $\Attb/\Al$ ratio
measured at the Tevatron as a function of the lepton $p_T$  
was shown to be a robust observable in the SM
in~\cite{Falkowski:2012cu}.
In other words, there is a correlation between
$\Attb$ and $\Al$ which qualitatively persists
from parton level to the level of including showering and reconstruction.
The  $A_{l}-A_{t\bar{t}}$ correlation shows stability under variations
of theoretical inputs and even under potential mismodeling.
The authors of~\cite{Falkowski:2012cu} also suggested the use of
$\mttb$ as an alternative kinematic variable to $\ptl$ but a concrete
study of its robustness was not provided. 
In this appendix we will show that the ratio of the Tevatron
asymmetries when measured as a function of $\mttb$ is stable against
the choice of renormalization and factorization scales and also
against a potential mis-modeling in the transverse momentum of the $t\bar{t}$ system.
Then, we will compare the recent D$\O$ measurement of the lepton asymmetry in the $l+$jets channel
as a function of lepton $p_{T}$ \cite{d06394} with the SM predictions from POWHEG~\cite{powheg} and MC@NLO~\cite{mcnlo}, this provides by itself a new robustness test for the correlation as explained below.
Finally, we give the results for the  differential $t \bar t$ forward-backward asymmetry $\Attb$ and the lepton-based forward-backward asymmetry $\Al$  as a function of $\mttb$ and $\ptl$ for the BSM benchmarks studied in this paper.

\subsection{Robustness Tests for the Differential Asymmetries}

In order to check the robustness of the $\Al/\Attb$ ratio when measured as a function of
$\mttb$, we have generated $t\bar{t}$ events with POWHEG, setting the renormalization and
factorization scales to $\mu_{R}=\mu_{F}=Q = \sqrt{m_{t}^{2}+(p_{T,t})^{2}}$ and using the MSTW2008NLO \cite{mstw} parton
distribution functions (this choice is made to match the choice of the parton distribution function made by the D$\O$ collaboration).
This study is performed at parton level without applying any kinematic cuts.

\begin{figure}[hbtp]
\centering
\includegraphics[width=7cm,height=5cm]{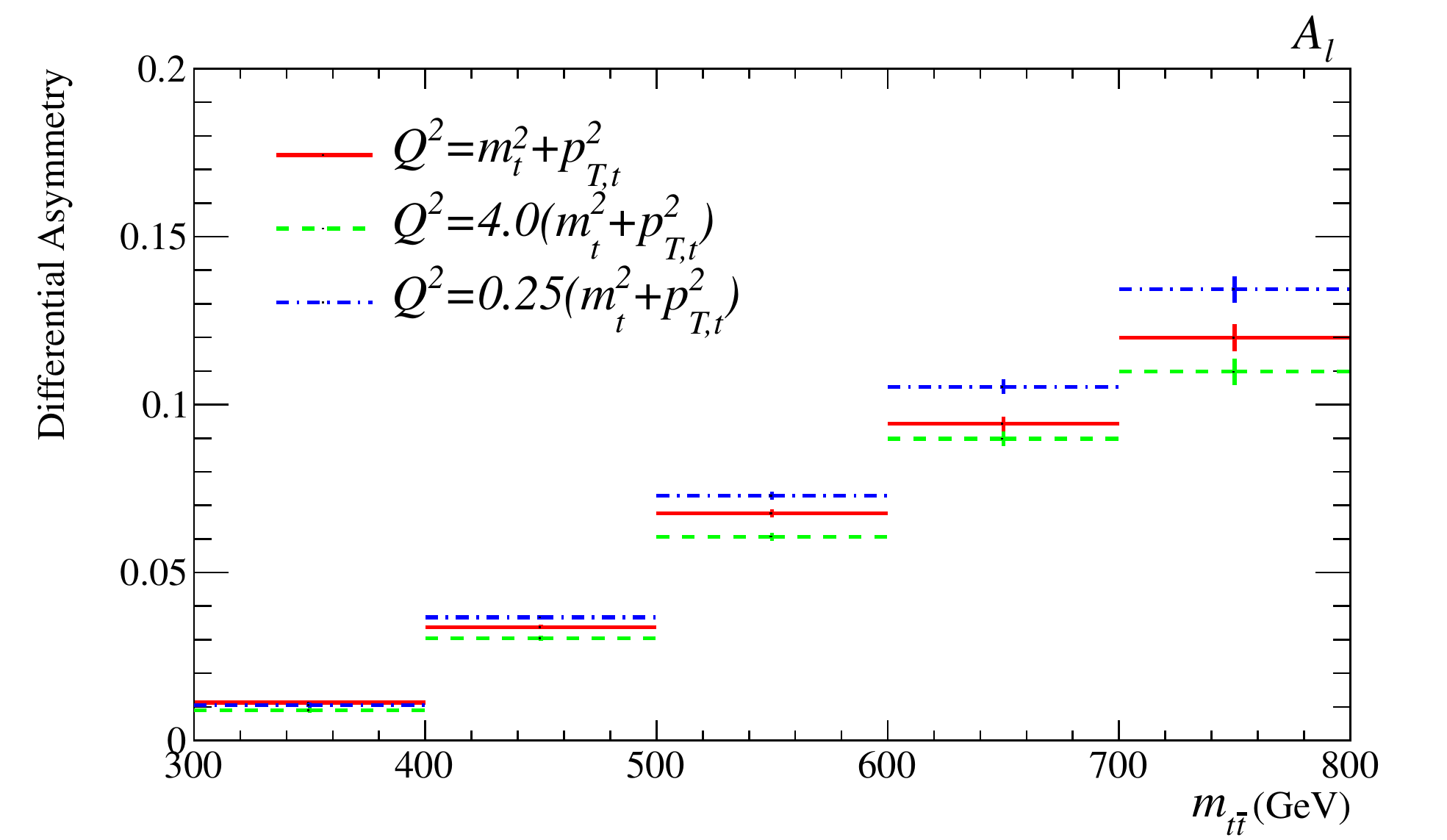}
\includegraphics[width=7cm,height=5cm]{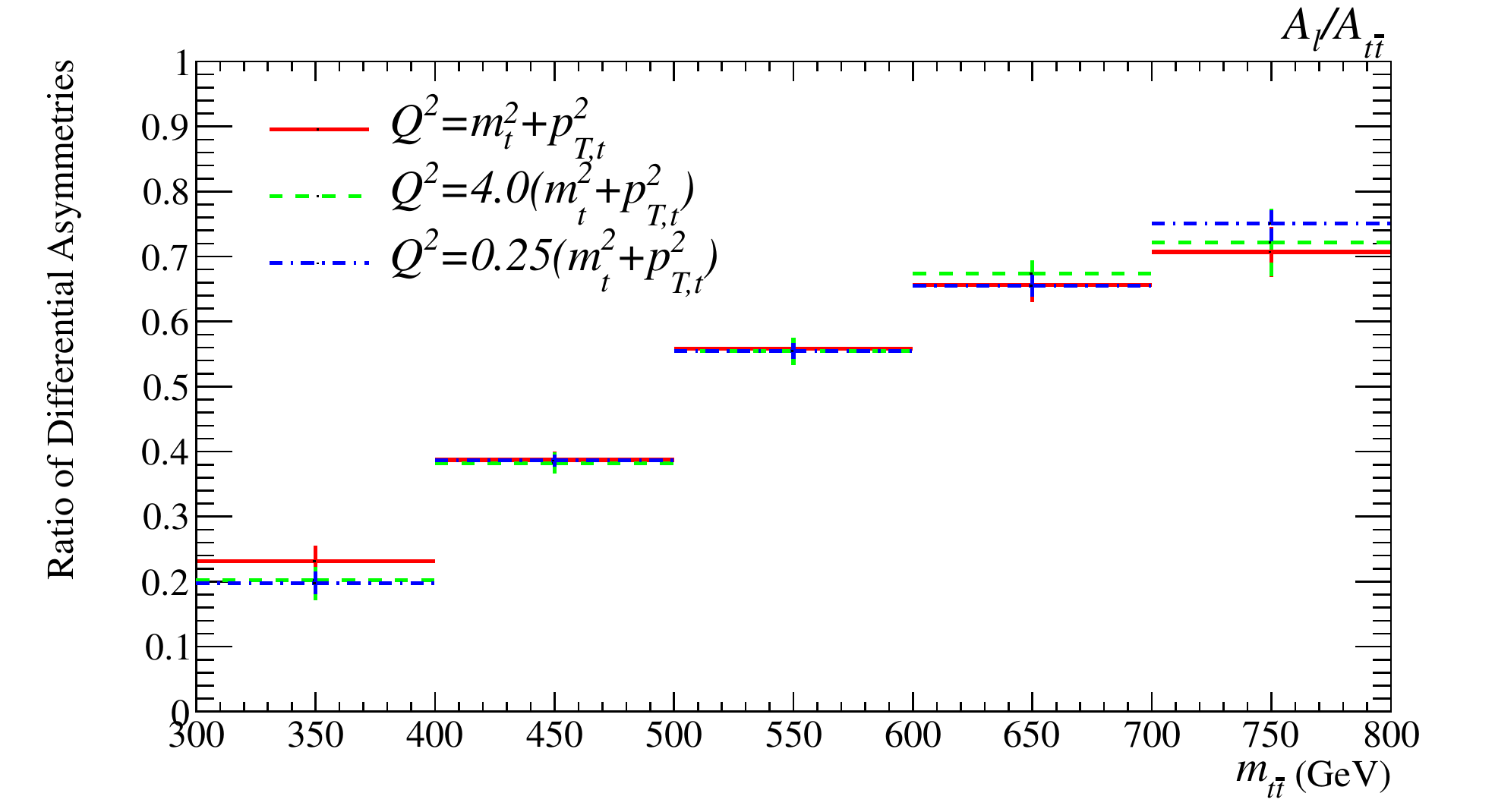}
\caption{Distribution of $\Al$ (left) and $\Al/\Attb$ (right) as a function of
  $\mttb$ for three different choices of the renormalization and
  factorization scale $Q$. These plots are for the ideal SM scenario with no cuts applied.}\label{tevQ}
\end{figure}

The impact of higher-order calculations can be estimated by varying
the renormalization and factorization scales. We have increased and reduced
the corresponding scales by a factor of two. We show in
Fig.~\ref{tevQ} the distribution of $\Al$ (left) and of the
$\Al/\Attb$ ratio (right) as a function of $\mttb$ for the three
choices of scales. As expected each individual asymmetry changes with
the scale but the ratio remains stable, showing that it is robust
against variations in the renormalization and factorization scales. 

\begin{figure}[hbtp]
\centering
\includegraphics[width=7cm,height=5cm]{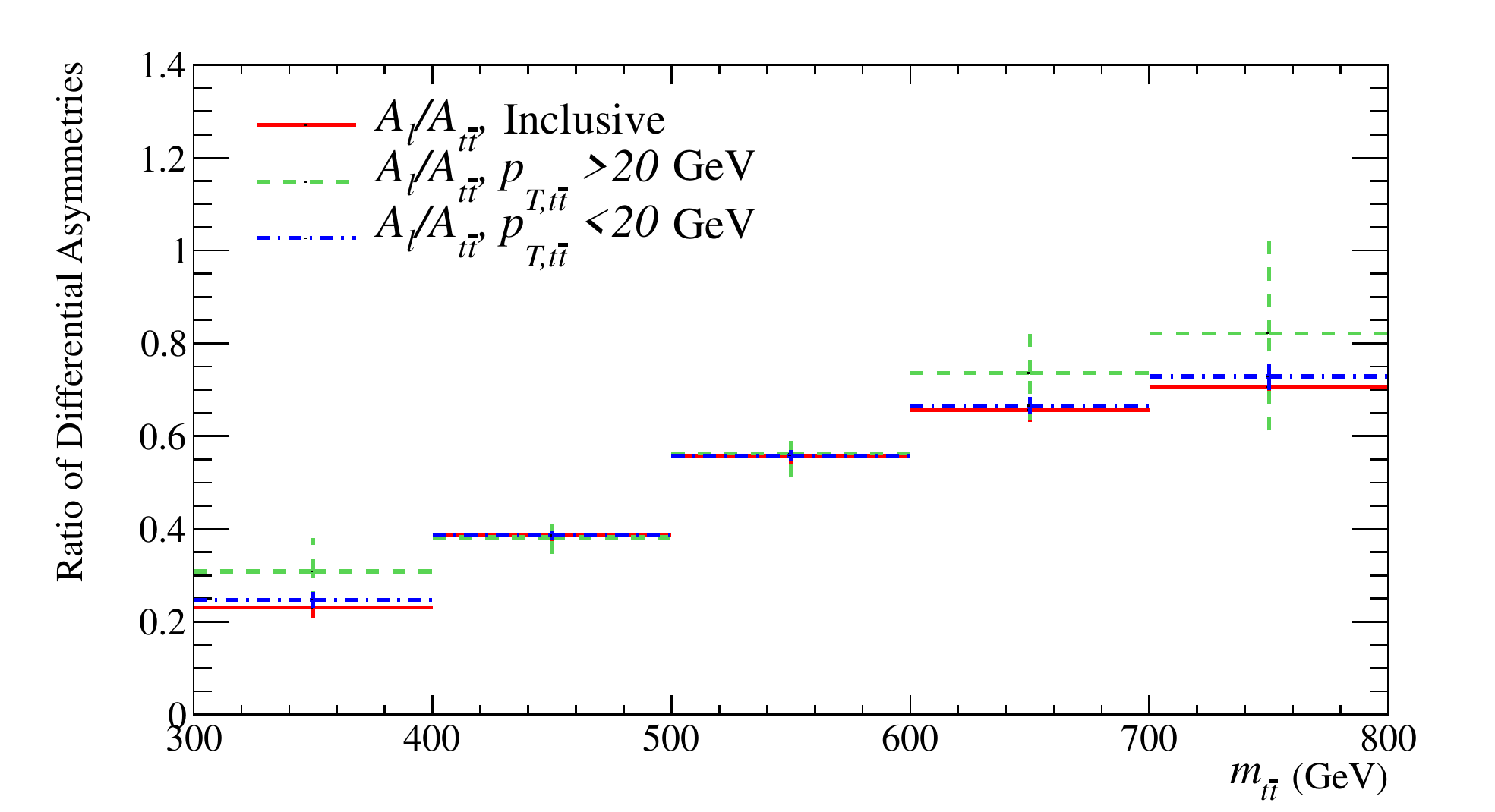}
\caption{The differential asymmetries ratio
 $A_{l}(m_{t\bar{t}})/A_{t\bar{t}}(m_{t\bar{t}})$ for two ranges
 of $p_{T,t\bar{t}}$ at the Tevatron.
 The blue and green depict the ratio for events with
 $p_{T,t\bar{t}}< 20$ GeV
and $p_{T,t\bar{t}} > 20$ GeV, respectively. The red curve is the
 ratio for all $p_{T,t\bar{t}}$ values.
 All the calculation are at NLO in idealized SM with the events
 simulated with POWHEG.}\label{tevPTT}
\end{figure}

It is known that the forward-backward asymmetry $A_{t\bar{t}}$ 
depends on the transverse  momentum of the $t\bar{t}$ system $p_{T,t\bar{t}}$ \cite{Bowen:2005ap}.
Therefore, another important robustness test is to verify the
sensitivity of correlation 
$A_{l}-A_{t\bar{t}}$ to the $p_{T,t\bar{t}}$.
In order to make sure that the correlation is not distorted in
different regions of $p_{T,t\bar{t}}$, we have calculated the ratio of
asymmetries in two separate $p_{T,\ttb}$ bins: $p_{T,t\bar{t}} < 20$ GeV
and  $p_{T,t\bar{t}} > 20$ GeV.
In Fig.~\ref{tevPTT}, we show the ratio
$\Al/\Attb$ as a function of $\mttb$ for $p_{T,t\bar{t}} <
20$ GeV, $p_{T,t\bar{t}} > 20$ GeV and for the inclusive case.
The result shows that the ratio is quite insensitive to the value of
$p_{T,\ttb}$ showing the robustness of the observable against 
the mismodelling of $p_{T,t\bar{t}}$.

\subsection{Lepton-based Asymmetry at the Tevatron versus Lepton $p_{T}$ and POWHEG versus MC@NLO}\label{Teva}

Based on the full Tevatron data sample  of 9.7 fb$^{-1}$, the  D$\O$ experiment has measured
the lepton forward-backward asymmetry in top pair events in the $l+$jets channel as a function of the lepton
transverse momentum \cite{d06394}.

 We now move to describe an additional new robustness test. We investigate whether the SM correlation between 
the lepton-based asymmetry versus $p_{T}$ of lepton is sensitive to the use of the specific tool that is used to calculated the matrix elements and parton shower.
For that purpose we compare
the POWHEG~\cite{powheg}
and MC@NLO~\cite{mcnlo} event generators.
These event generators are suitable for this measurement since both include
the NLO calculation of top pair production with
subsequent simulation of parton showers.
The $t\bar{t}$ events generated with MC@NLO have been combined with
HERWIG~\cite{herwig}   for showering and hadronization 
and the POWHEG events have been combined with PYTHIA~\cite{pythia} for parton showering and
hadronization. 
The outputs have been passed through FASTJET \cite{fastjet} to
reconstruct the jets. 
After this step, we have applied similar cuts and requirements as in
\cite{d06394}. 

\begin{figure}[hbtp]
\centering
\includegraphics[width=7cm,height=5cm]{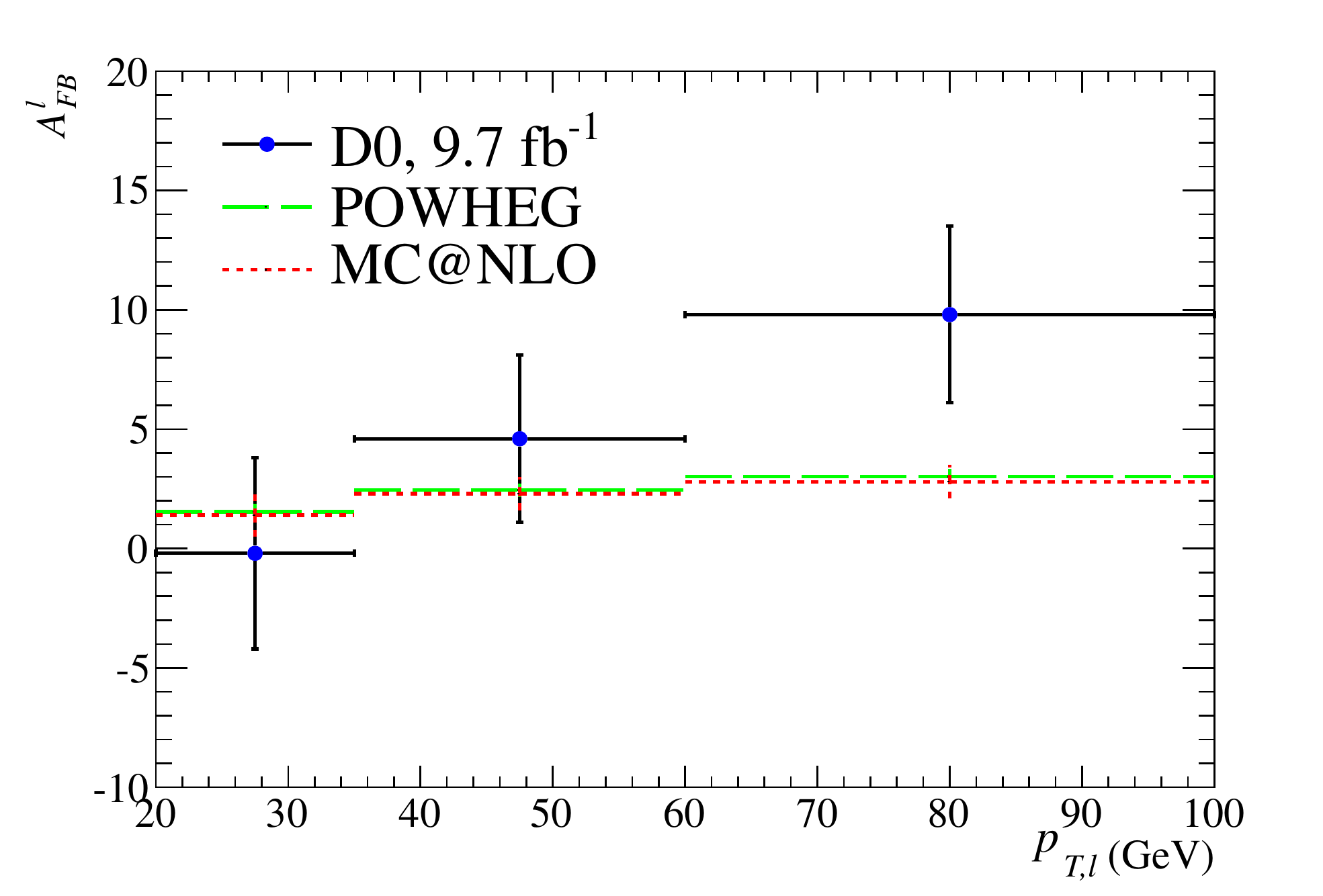}
\caption{The observed and predictions of the lepton-based asymmetries
  as a function of the charged lepton transverse momentum.
The green curve has been obtained with POWHEG and the blue is the output
  of MC@NLO. } \label{Ald0}
\end{figure}

We have then computed the lepton asymmetry defined in
Eq. (\ref{eq:Al:def}) 
in the following three $\ptl$ bins:
$20 < p_{T} < 35$ GeV (low),
$35 < p_{T} < 60$ GeV (mid) and $p_{T} > 60$ GeV (high).
The measurements from D$\O$ and the SM prediction obtained by us using
MC@NLO and POWHEG are reported in Table \ref{AlTev} and plotted in
Fig.\ref{Ald0} for comparison. 
As can be seen, the asymmetries computed with MC@NLO and POWHEG are
virtually identical in all three bins. They are compatible with the
D$\O$ measurements in the first two bins and show a slight excess in
the largest $\ptl$ bin. Clearly, a measurement of the $\Al/\Attb$
ratio as a function of $\ptl$ could provide a very valuable
information on the possible origin of this excess.

\begin{table}\caption{The SM predicted values and the observed lepton asymmetries in three bins of lepton $p_{T}$.}
\begin{center}
\begin{tabular}{c|cccc}
  $A_{l}\%$ & Inclusive & Low $p_{T}$ & Mid $p_{T}$ & High $p_{T}$ \\ \hline
  Data   &\rule{0mm}{5mm} $4.7 \pm 2.3^{+1.1}_{-1.4}$ & $-0.2 \pm 4.0^{+1.7}_{-2.3}$ & $4.6 \pm 3.5^{+1.8}_{-1.3}$ & $9.8 \pm 3.7^{+1.9}_{-2.2}$ \\
  MC@NLO & $2.2\pm 0.5$ & $1.4 \pm 0.9 $ & $2.3 \pm 0.7$ & $2.8 \pm 0.7$ \\
  POWHEG & $2.41\pm 0.18$ & $1.54 \pm 0.33$ & $2.54 \pm 0.28$ & $3.02 \pm 0.31$
\end{tabular}\label{AlTev}
\end{center}
\end{table}

\subsection{BSM benchmarks}

\begin{figure}[hbtp]
\centering
\includegraphics[width=0.45\textwidth]{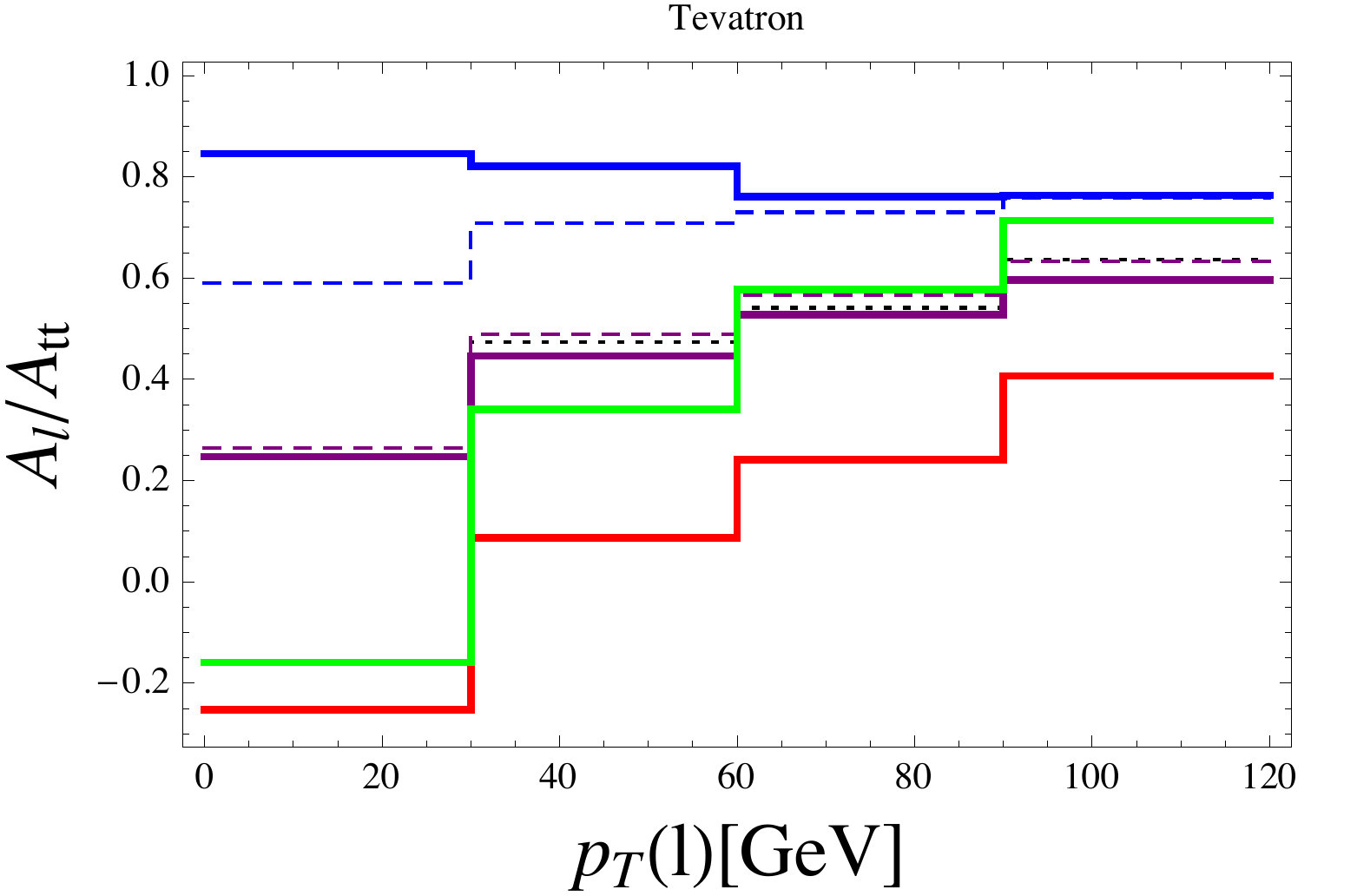}
\includegraphics[width=0.45\textwidth]{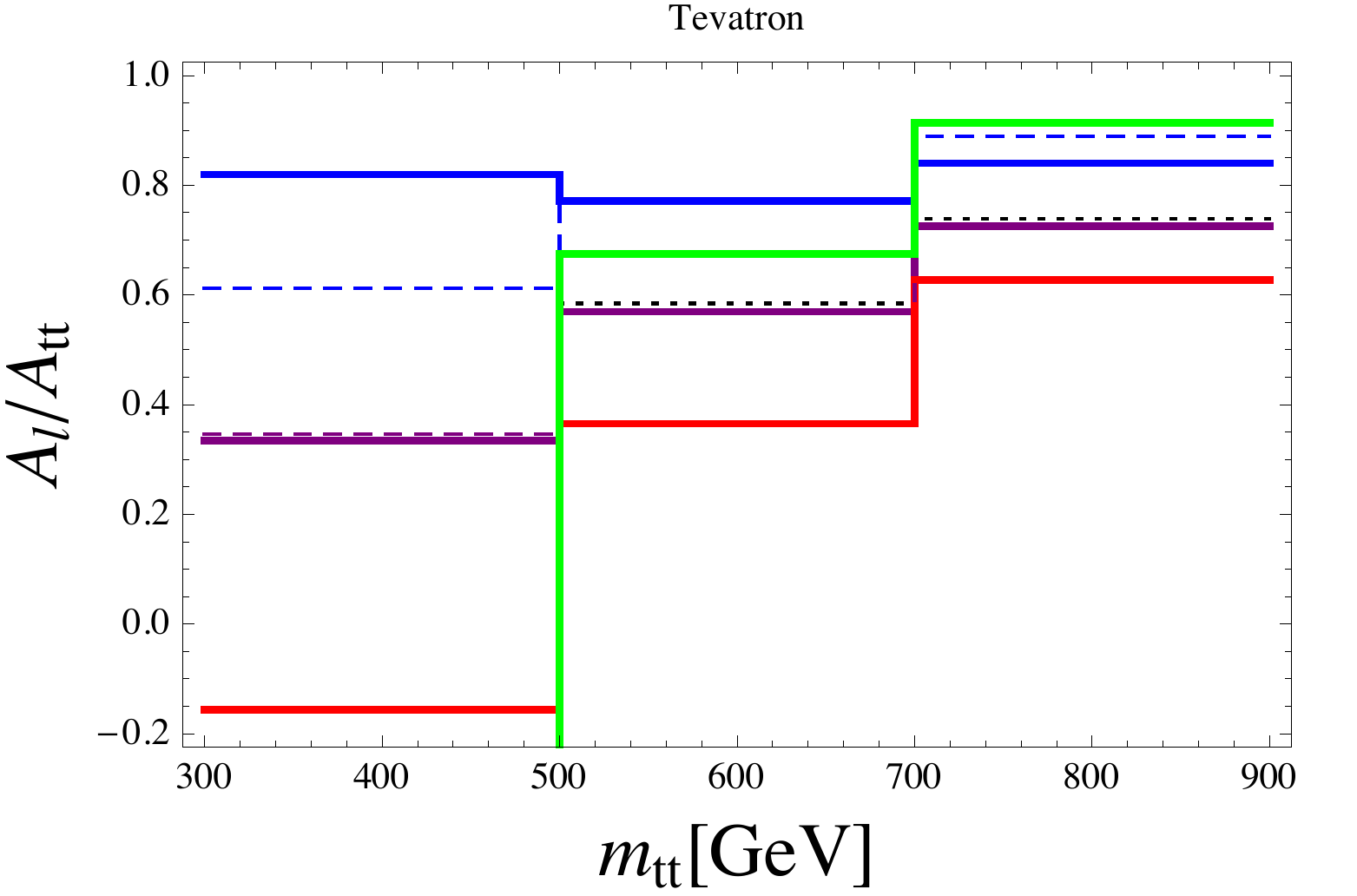}
\caption{Distribution of the ratio $ \Al/\Attb$ at the Tevatron as  a  function of $\ptl$ (left) and $\mttb$ (right) for the SM (dotted
  black)  and for the BSM benchmarks studied in this paper:  Axi200R (solid blue), Axi200L (solid red), Axi200A (solid purple), Axi2000R (dashed blue), Axi2000A (dashed purple), and Zp220 (solid green).     
 }
 \label{fig:rbsmtev}
\end{figure}

For completeness, in Tables~\ref{tab:attbsm}~and~\ref{tab:albsm}  we list the differential results for the Tevatron $t \bar t$ forward-backward asymmetry $\Attb$ and the lepton-based forward-backward asymmetry $\Al$  as a function of $\mttb$ and $\ptl$ for the BSM benchmarks studied in this paper. 
In Fig.~\ref{fig:rbsmtev} we plot the ratio of these differential asymmetries.   
As in the case of the LHC, the ratio has a strong discriminating power for BSM models where  $t \bar t$  production is polarized.  

\begin{table} 
\begin{tabular}{l|c|c|c|c|}
$\ptl$[GeV] & [0,30]  &  [30,60] &  [60,90] & [90,120]    
 \\ \hline 
{\bf Axi200R }&  0.11 & 0.11 & 0.12 & 0.13 
 \\  \hline  
{\bf Axi200L }&  0.13 & 0.11  & 0.19 & 0.10 
\\ \hline 
{\bf  Axi200A }& 0.21 & 0.20 &  0.19 & 0.20 
\\  \hline 
{\bf  Axi2000R }&  0.11 & 0.10 & 0.11 & 0.13  
\\  \hline 
{\bf  Axi2000A }& 0.15 & 0.14 & 0.15 & 0.17  
\\ \hline 
{\bf Zp220 } & 0.21 & 0.22 & 0.24  & 0.27  
\end{tabular}
\quad 
\begin{tabular}{l|c|c|c|c|}
$\mttb$[GeV]  & [300,500] &  [500,700] &  [700,900]   
 \\ \hline 
{\bf Axi200R }& 0.10 & 0.17 & 0.23 
 \\  \hline  
{\bf Axi200L }&  0.10 & 0.18  & 0.23  
\\ \hline 
{\bf  Axi200A }& 0.17 & 0.30 &  0.37  
\\  \hline 
{\bf  Axi2000R }& 0.08 & 0.18 & 0.30  
\\  \hline 
{\bf  Axi2000A }& 0.11 & 0.27 & 0.49 
\\ \hline 
{\bf Zp220 } & 0.10  & 0.43 & 0.74
\end{tabular}
\caption{$\Attb$ at the Tevatron as a function of $\ptl$ (left) and $\mttb$ (right) for the benchmarks studied in this paper. }
\label{tab:attbsm}
\end{table}

\begin{table} 
\begin{tabular}{l|c|c|c|c|}
$\ptl$[GeV]  & [0,30] &  [30,60] &  [60,90] & [90,120]    
 \\ \hline 
{\bf Axi200R }& 0.10 & 0.09 & 0.09 & 0.10 
 \\  \hline  
{\bf Axi200L }& -0.03 & 0.01 & 0.03 & 0.04 
\\ \hline 
{\bf  Axi200A }& 0.05 & 0.09 &  0.10 & 0.12 
\\  \hline 
{\bf  Axi2000R }&  0.06 & 0.07 & 0.08 & 0.10  
\\  \hline 
{\bf  Axi2000A }& 0.04  & 0.07 & 0.08 & 0.11  
\\ \hline 
{\bf Zp220 } & -0.03 & 0.07 & 0.14  & 0.19  
\end{tabular}
\quad
\begin{tabular}{l|c|c|c|c|}
$\mttb$[GeV]  & [300,500] &  [500,700] &  [700,900]   
 \\ \hline 
{\bf Axi200R }& 0.08 & 0.13 & 0.19  
 \\  \hline  
{\bf Axi200L }& -0.02 & 0.06 & 0.15  
\\ \hline 
{\bf  Axi200A }& 0.06 & 0.17 &  0.27  
\\  \hline 
{\bf  Axi2000R }&  0.05 & 0.14 & 0.27  
\\  \hline 
{\bf  Axi2000A }& 0.04 & 0.15 & 0.36  
\\ \hline 
{\bf Zp220 } & -0.07 & 0.29 & 0.67 
\end{tabular}

\caption{$\Al$ at the Tevatron as a function of $\ptl$ (left) and $\mttb$ (right)  for the benchmarks studied in this paper.} 
\label{tab:albsm}
\end{table}

\begin{thebibliography}{}

\bibitem{Carena:2007ua} 
  M.~S.~Carena, E.~Ponton, J.~Santiago and C.~E.~M.~Wagner,
  Phys.\ Rev.\ D {\bf 76}, 035006 (2007)
  [hep-ph/0701055].
  

  
\bibitem{Atre:2008iu} 
  A.~Atre, M.~Carena, T.~Han and J.~Santiago,
  Phys.\ Rev.\ D {\bf 79}, 054018 (2009)
  [arXiv:0806.3966 [hep-ph]].


\bibitem{flavor-triviality1}
  C.~Delaunay, O.~Gedalia, S.~J.~Lee, G.~Perez and E.~Ponton,
  Phys.\ Rev.\ D {\bf 83}, 115003  (2011) 
  [arXiv:1007.0243 [hep-ph]];
   C.~Delaunay, O.~Gedalia, S.~J.~Lee, G.~Perez adnd E.~Ponton,
  Phys.\ Lett.\ B {\bf 703}, 486 (2011)
  [arXiv:1101.2902 [hep-ph]].
  
\bibitem{Redi:2011zi}
  M.~Redi and A.~Weiler,
  JHEP {\bf 1111}, 108  (2011)
  [arXiv:1106.6357 [hep-ph]].
  
\bibitem{Atre:2013ap} 
  A.~Atre, M.~Chala and J.~Santiago,
  JHEP {\bf 1305}, 099 (2013)
  [arXiv:1302.0270 [hep-ph]].
  
\bibitem{Delaunay:2013iia}
  C.~Delaunay, C.~Grojean and G.~Perez,
  arXiv:1303.5701 [hep-ph].


\bibitem{DaRold:2012sz}
  L.~Da Rold, C.~Delaunay, C.~Grojean and G.~Perez,
  JHEP {\bf 1302}, 149  (2013) 
  [arXiv:1208.1499 [hep-ph]].


\bibitem{Kuhn:1998kw}
  J.~H.~Kuhn and G.~Rodrigo,
  Phys.\ Rev.\ D {\bf 59} (1999) 054017
  [hep-ph/9807420].
\bibitem{Kuhn:1998jr}
  J.~H.~Kuhn and G.~Rodrigo,
  Phys.\ Rev.\ Lett.\  {\bf 81} (1998) 49
  [hep-ph/9802268].

\bibitem{Bowen:2005ap}
  M.~T.~Bowen, S.~D.~Ellis and D.~Rainwater,
  Phys.\ Rev.\ D {\bf 73} (2006) 014008
  [hep-ph/0509267].
  
  \bibitem{Antunano:2007da}
  O.~Antunano, J.~H.~Kuhn and G.~Rodrigo,
  Phys.\ Rev.\ D {\bf 77} (2008) 014003
  [arXiv:0709.1652 [hep-ph]].

  
\bibitem{Almeida:2008ug}
  L.~G.~Almeida, G.~F.~Sterman and W.~Vogelsang,
  Phys.\ Rev.\ D {\bf 78} (2008) 014008
  [arXiv:0805.1885 [hep-ph]].




\bibitem{Aaltonen:2012it}
  T.~Aaltonen {\it et al.}  [CDF Collaboration],
  Phys.\ Rev.\ D {\bf 87}, 092002 (2013)
  [arXiv:1211.1003 [hep-ex]].


\bibitem{Abazov:2011rq}
  V.~M.~Abazov {\it et al.}  [D0 Collaboration],
  Phys.\ Rev.\ D {\bf 84}, 112005 (2011)
  [arXiv:1107.4995 [hep-ex]].


\bibitem{Bernreuther:2012sx} 
  W.~Bernreuther and Z.~-G.~Si,
  Phys.\ Rev.\ D {\bf 86}, 034026 (2012)
  [arXiv:1205.6580 [hep-ph]].


\bibitem{Aaltonen:2013vaf}
T.~A.~Aaltonen {\it et al.} [CDF Collaboration],
Phys.\ Rev.\ D {\bf 88} (2013) 072003
[arXiv:1308.1120 [hep-ex]].


\bibitem{D0:Alsemi-leptonic}
D0 Note Conf 6381. 

\bibitem{cdfall}  CDF Note 11035. 

\bibitem{Abazov:2013wxa}
  V.~M.~Abazov {\it et al.}  [D0 Collaboration],
  arXiv:1308.6690 [hep-ex].





\bibitem{Falkowski:2012cu}
A.~Falkowski, M.~L.~Mangano, A.~Martin, G.~Perez and J.~Winter,
Phys.\ Rev.\ D {\bf 87} (2013) 034039
[arXiv:1212.4003 [hep-ph]].

\bibitem{Falkowski:2011zr} 
  A.~Falkowski, G.~Perez and M.~Schmaltz,
  Phys.\ Rev.\ D {\bf 87}, 034041 (2013)
  [arXiv:1110.3796 [hep-ph]].
  
  
\bibitem{Berger:2011pu} 
  E.~L.~Berger, Q.~-H.~Cao, C.~-R.~Chen, J.~-H.~Yu and H.~Zhang,
  arXiv:1111.3641 [hep-ph].
  E.~L.~Berger, Q.~-H.~Cao, C.~-R.~Chen, J.~-H.~Yu and H.~Zhang,
  Phys.\ Rev.\ Lett.\  {\bf 108}, 072002 (2012)
  [arXiv:1201.1790 [hep-ph]].
  E.~L.~Berger, Q.~-H.~Cao, C.~-R.~Chen and H.~Zhang,
  Phys.\ Rev.\ D {\bf 88}, 014033 (2013)
  [arXiv:1209.4899 [hep-ph]].
  
\bibitem{Baumgart:2013yra} 
  M.~Baumgart and B.~Tweedie,
  JHEP {\bf 1308}, 072 (2013)
  [arXiv:1303.1200 [hep-ph]].

\bibitem{Agashe:2006hk} K.~Agashe, A.~Belyaev, T.~Krupovnickas, G.~Perez and J.~Virzi,
  Phys.\ Rev.\ D {\bf 77}, 015003 (2008)
  [hep-ph/0612015];


\bibitem{Drobnak:2012cz}
  J.~Drobnak, J.~F.~Kamenik and J.~Zupan,
  Phys.\ Rev.\ D {\bf 86} (2012) 054022
  [arXiv:1205.4721 [hep-ph]].
  
\bibitem{Drobnak:2012rb}
J.~Drobnak, A.~L.~Kagan, J.~F.~Kamenik, G.~Perez and J.~Zupan,
Phys.\ Rev.\ D {\bf 86} (2012) 094040
[arXiv:1209.4872 [hep-ph]].

  
\bibitem{AguilarSaavedra:2012va}
  J.~A.~Aguilar-Saavedra and A.~Juste,
  Phys.\ Rev.\ Lett.\  {\bf 109} (2012) 211804
  [arXiv:1205.1898 [hep-ph]].



\bibitem{Alvarez:2012ca}
  E.~Alvarez and E.~C.~Leskow,
  Phys.\ Rev.\ D {\bf 86} (2012) 114034
  [arXiv:1209.4354 [hep-ph]].
  

\bibitem{A_C:semi-leptonic:atlas}
ATLAS conference note ATLAS-CONF-2013-078.

\bibitem{Chatrchyan:2012cxa}
S.~Chatrchyan {\it et al.} [CMS Collaboration],
Phys.\ Lett.\ B {\bf 717} (2012) 129
[arXiv:1207.0065 [hep-ex]].

\bibitem{A_C:semi-leptonic:cms}
CMS conference note CMS PAS TOP-12-033.

\bibitem{A_C:leptonic:atlas}
ATLAS conference note ATLAS-CONF-2012-057.
\bibitem{A_C:leptonic:cms}
CMS conference note CMS PAS TOP-12-010.

\bibitem{powheg}
 S.~Alioli, P.~Nason, C.~Oleari and E.~Re,
  JHEP {\bf 1006}, 043 (2010)
  [arXiv:1002.2581 [hep-ph]].

\bibitem{pdfct10}
  H.~-L.~Lai, M.~Guzzi, J.~Huston, Z.~Li, P.~M.~Nadolsky, J.~Pumplin and C.~-P.~Yuan,
  Phys.\ Rev.\ D {\bf 82}, 074024 (2010)
  [arXiv:1007.2241 [hep-ph]].


\bibitem{acpaper}
  J.~H.~Kuhn and G.~Rodrigo,
  JHEP {\bf 1201}, 063 (2012)
  [arXiv:1109.6830 [hep-ph]].

\bibitem{mcnlo}
  S.~Frixione and B.~R.~Webber,
  JHEP {\bf 0206}, 029 (2002)
  [hep-ph/0204244];
  S.~Frixione, P.~Nason and B.~R.~Webber,
  JHEP {\bf 0308}, 007 (2003)
  [hep-ph/0305252].




\bibitem{herwig}
  G.~Corcella, I.~G.~Knowles, G.~Marchesini, S.~Moretti, K.~Odagiri, P.~Richardson, M.~H.~Seymour and B.~R.~Webber,
  JHEP {\bf 0101}, 010 (2001)
  [hep-ph/0011363].

\bibitem{pythia}
 T.~Sjostrand, L.~Lonnblad, S.~Mrenna and P.~Z.~Skands,
  hep-ph/0308153.


\bibitem{AguilarSaavedra:2011hz} 
  J.~A.~Aguilar-Saavedra and M.~Perez-Victoria,
  Phys.\ Rev.\ D {\bf 84}, 115013 (2011)
  [arXiv:1105.4606 [hep-ph]].

\bibitem{AguilarSaavedra:2011ug} 
  J.~A.~Aguilar-Saavedra and M.~Perez-Victoria,
  JHEP {\bf 1109}, 097 (2011)
  [arXiv:1107.0841 [hep-ph]].

\bibitem{Fajfer:2012si} 
  S.~Fajfer, J.~F.~Kamenik and B.~Melic,
  JHEP {\bf 1208}, 114 (2012)
  [arXiv:1205.0264 [hep-ph]].

\bibitem{Aaltonen:2013wca} 
  T.~A.~Aaltonen {\it et al.}  [ CDF and  D0 Collaborations],
  arXiv:1309.7570 [hep-ex].

\bibitem{Czakon:2013goa}
M.~Czakon, P.~Fiedler and A.~Mitov,
Phys.\ Rev.\ Lett.\ {\bf 110} (2013) 252004
[arXiv:1303.6254 [hep-ph]].




\bibitem{Aaltonen:2009iz}
T.~Aaltonen {\it et al.} [CDF Collaboration],
Phys.\ Rev.\ Lett.\ {\bf 102} (2009) 222003
[arXiv:0903.2850 [hep-ex]].


\bibitem{D0mtt} D$\O$ Conference note 6379. 

\bibitem{Chatrchyan:2013lca}
S.~Chatrchyan {\it et al.} [CMS Collaboration],
arXiv:1309.2030 [hep-ex].



\bibitem{Hall:1985wz}
  L.~J.~Hall and A.~E.~Nelson,
  Phys.\ Lett.\ B {\bf 153}, 430 (1985).
  P.~H.~Frampton and S.~L.~Glashow,
  Phys.\ Lett.\ B {\bf 190}, 157 (1987).


\bibitem{Tavares:2011zg}
G.~Marques Tavares and M.~Schmaltz,
Phys.\ Rev.\ D {\bf 84} (2011) 054008
[arXiv:1107.0978 [hep-ph]].



\bibitem{AguilarSaavedra:2011ci}
J.~A.~Aguilar-Saavedra and M.~Perez-Victoria,
Phys.\ Lett.\ B {\bf 705} (2011) 228
[arXiv:1107.2120 [hep-ph]].

\bibitem{AguilarSaavedra:2011ck} 
  J.~A.~Aguilar-Saavedra and J.~Santiago,
  Phys.\ Rev.\ D {\bf 85}, 034021 (2012)
  [arXiv:1112.3778 [hep-ph]].

\bibitem{Gross:2012bz}
C.~Gross, G.~Marques Tavares, M.~Schmaltz and C.~Spethmann,
Phys.\ Rev.\ D {\bf 87} (2013) 014004
[arXiv:1209.6375 [hep-ph]].

\bibitem{Gresham:2012kv} 
  M.~Gresham, J.~Shelton and K.~M.~Zurek,
  JHEP {\bf 1303}, 008 (2013)
  [arXiv:1212.1718 [hep-ph]].


\bibitem{Barcelo:2011vk} 
  R.~Barcelo, A.~Carmona, M.~Masip and J.~Santiago,
  Phys.\ Lett.\ B {\bf 707}, 88 (2012)
  [arXiv:1106.4054 [hep-ph]].


 
\bibitem{Jung:2011zv} 
  S.~Jung, A.~Pierce and J.~D.~Wells,
  Phys.\ Rev.\ D {\bf 83}, 114039 (2011)
  [arXiv:1103.4835 [hep-ph]].


\bibitem{Jung:2009jz} 
  S.~Jung, H.~Murayama, A.~Pierce and J.~D.~Wells,
  Phys.\ Rev.\ D {\bf 81}, 015004 (2010)
  [arXiv:0907.4112 [hep-ph]].
  
\bibitem{AguilarSaavedra:2011zy} 
  J.~A.~Aguilar-Saavedra and M.~Perez-Victoria,
  Phys.\ Lett.\ B {\bf 701}, 93 (2011)
  [arXiv:1104.1385 [hep-ph]].

\bibitem{Alwall:2011uj} 
  J.~Alwall, M.~Herquet, F.~Maltoni, O.~Mattelaer and T.~Stelzer,
  JHEP {\bf 1106}, 128 (2011)
  [arXiv:1106.0522 [hep-ph]].

\bibitem{d06394}
D$\O$ Conference note 6394.


\bibitem{mstw}
  A.~D.~Martin, W.~J.~Stirling, R.~S.~Thorne and G.~Watt,
  Eur.\ Phys.\ J.\ C {\bf 63}, 189 (2009)
  [arXiv:0901.0002 [hep-ph]].


\bibitem{fastjet}
M.~Cacciari, G.~P.~Salam and G.~Soyez,
  Eur.\ Phys.\ J.\ C {\bf 72}, 1896 (2012)  [arXiv:1111.6097 [hep-ph]].













  
\end{thebibliography}
\end{document}